	\documentclass[11pt, draftclsnofoot, onecolumn]{IEEEtran}
	\usepackage{pifont,bm,multicol,amsfonts,amsmath,color,amssymb,graphicx, epsfig,cite,psfrag,subfigure,algorithm,enumerate,stfloats,algorithm,algorithmic,epstopdf,balance}

	\newtheorem{lemma}{\underline{Lemma}}

	\def\tr{\mathop\mathrm{tr}}

\begin{document}
\title{Time-Varying Massive MIMO Channel Estimation: Capturing, Reconstruction and Restoration}

\author{Muye Li,  Shun Zhang,{ \emph{Member, IEEE,}} Nan Zhao, \emph{Senior Member, IEEE,} \\   Weile Zhang,{ \emph{Member, IEEE,}} Xianbin Wang, \emph{Fellow, IEEE}
	
    \thanks{M. Li, S. Zhang are with the State Key Laboratory of Integrated Services Networks, Xidian University, Xi’an 710071, P. R. China (Email: limuyexdu@163.com; zhangshunsdu@xidian.edu.cn).}

    \thanks{N. Zhao is with the School of Information and Communication Engineering, Dalian University of Technology, Dalian 116024, P. R. China (zhaonan@dlut.edu.cn).}

    \thanks{W. Zhang is with the MOE Key Lab for Intelligent Networks and Network Security, Xi'an Jiaotong University, Xi'an 710049, P. R. China (Email: wlzhang@mail.xjtu.edu.cn).}

    \thanks{X. Wang is with Department of Electrical and Computer Engineering, Western University, London, Ontario, Canada (Email: xianbin.wang@uwo.ca).}
}

\maketitle


\vspace{-10mm}
\begin{abstract}
On {the} time-varying channel estimation,
the traditional downlink (DL) channel restoration schemes usually require the reconstruction for the covariance of downlink process noise vector,
which is dependent on DL channel covariance matrix (CCM).
However, the acquisition of the CCM leads to unacceptable overhead in massive MIMO systems.
To tackle this problem, in this paper,
we propose a {novel} scheme for the DL channel tracking.
First, {with the help of virtual channel representation (VCR)}, we build a dynamic uplink (UL) massive MIMO channel model with the consideration of off-grid refinement.
Then, a coordinate-wise maximization based expectation maximization (EM) algorithm is adopted for {capturing} the model parameters, {including the spatial signatures, the time-correlation factors, the off-grid bias, the channel power, and the noise power.}
Thanks to the angle reciprocity, {the spatial signatures, time-correlation factors and off-grid bias} of the DL channel model
can be {reconstructed} with the knowledge of UL ones.
{However, the other two kinds of  model parameters are closely related with the carrier frequency,
which cannot be perfectly inferred from the UL ones. }
{Instead of relearning the DL model parameters with dedicated training},
we resort to the optimal Bayesian Kalman filter (OBKF) method
to accurately track the DL channel with the partially prior knowledge.
{At the same time, the model parameters will be gradually  restored.}
Specially, the factor-graph and the Metropolis Hastings MCMC are utilized {within the OBKF framework}.
%As our method do not require the knowledge of channel process noise covariance as well as the transmission noise covariance,
%the feedback overhead will be eliminated, while the tracking accuracy is also acceptable.
Finally, numerical results are provided to demonstrate the {efficiency} of our proposed scheme.

\end{abstract}

\maketitle \thispagestyle{empty} \vspace{-1mm}


\begin{IEEEkeywords}
	Massive MIMO,  sparse Bayesian learning, time-varying channels, {optimal} Bayesian Kalman filter, factor graph
\end{IEEEkeywords}

% \fi
\IEEEpeerreviewmaketitle

\section{Introduction}
Due to its tremendous improvement in {the} spectral and energy efficiency \cite{massive_MIMO},
massive multiple-input multiple-output (MIMO) has become a potential technology for the 5th generation (5G) cellular networks
to meet the future capacity requirement\cite{efficiency2,Massive_in_5G_1, Hai_Lin_JSAC, JIN_dft}.
In order to exploit the advantages of massive MIMO, perfect channel state information (CSI) is indispensable at the base station (BS).
In {the} time-division duplex (TDD) systems, as there exists reciprocity between the uplink (UL) and downlink (DL) channel,
the CSI at BS side {can be} obtained through UL training\cite{feedback,yang2016design}.
However, in {the} frequency-division duplex (FDD) systems, the CSI at BS side should be obtained through the uplink training, downlink training and CSI feedback,
which will cause unaffordable overhead together with the pilot contamination \cite{training_fdd2, lowresolutionADC, gao_access, pilotcontamination}.

Recently, to reduce the overhead of channel training and the CSI feedback,
a set of new transmission strategies were proposed
{to reduce} the dimensions of the effective channels, {where low-rank property of the massive MIMO channel covariance matrix was fully exploited}\cite{Xie2018CCM, JSDM,JSDM_Opportunistic,JSDM_mm,beam_division,cov_estimation_FDD}.
In \cite{JSDM}, a joint spatial division multiplexing (JSDM) scheme was proposed
{to} project the eigenspace of channel covariance matrix of the desired user
into the nullspace of the eigenspaces {for} all other users and {to force} the inter-user interference to zero.
Nam et al. extended the results in \cite{JSDM},
and designed a low-cost opportunistic user selection and prebeamforming algorithm to achieve the optimal sum-rate in \cite{JSDM_Opportunistic} .
In \cite{JSDM_mm}, {Adhikary \emph{et al.} improved the JSDM scheme {to} decrease the computational complexity.}
Sun et al. proposed a complete transmission scheme named beam division multiplex for FDD massive MIMO system under two-stage precoding framework in \cite{beam_division},
where only the statistical CSI is used for the optimal downlink transmission.
Xie et al. proposed a new channel estimation scheme for TDD/FDD massive MIMO system \cite{Xie2018CCM},
{where the UL/DL channel covariance matrices (CCM) were reconstructed.}
{In this paper,} the authors extracted the angle parameters and power angular spectrum (PAS) of channel from the instantaneous uplink CSI,
{reconstructed the UL CCM} and used it to improve the UL channel estimation without any additional training cost,
which does not need the long-time acquisition for uplink CCMs
and can handle a more practical channel propagation environment with larger AS.
All the above methods utilize the spatial information for the implementation of orthogonal transmission {to} different users.
Theoretically, the spatial information can be derived from channel covariance matrix.
Thus, the low-complex and effective achieving methods for the channel covariance matrices are significant to
the above works \cite{JSDM,JSDM_Opportunistic,JSDM_mm,beam_division, Xie2018CCM,cov_estimation_FDD}.

Nevertheless, it is quite difficult to acquire channel covariances in massive MIMO system \cite{IASSR},
{which is due to the singular value decomposition (SVD) of the high-dimensional matrix.}
To overcome the bottleneck,
Xie et al. built a low-rank model for the instantaneous massive MIMO channel from antenna array theory \cite {gao_E} and proposed a spatial basis expansion model (SBEM)
to offer an alternative for the channel acquisition without the channel covariances.
Tang et al. proposed an off-grid channel estimation algorithm for the UL millimeter wave massive MIMO systems \cite{offgridSBL}.
The authers exploited the physical structure of CSI and proposed an {improved sparse Bayesian learning (ISBL)} algorithm which can achieve high estimation accuracy.
In \cite{time-vary-gao1}, a new channel tracking method for massive MIMO systems was proposed under {the time-varying} circumstance.
The extended KF method was used to blindly track the central angle, and the Taylor series expansion of the steering vectors was adopted to obtain the angular spread.
{Our previous work} \cite{Ma_J_SBL_Time_varing} proposed a channel estimation scheme for {the time-varying massive MIMO networks}.
{The} authors developed a EM-based SBL framework to learn {the temporal correlation factor, spatial signatures, and the channel powers,}
while Kalman filter (KF) and Rauch-Tung-Striebel smoother (RTSS) were adopted.
Then they applied a reduced dimension KF for UL/DL virtual channel tracking.
{However, the channel powers are closely related with the carrier frequency, which can not be perfectly inferred from the UL ones. Moreover, considering
the randomness of direction of arrivals (DOAs) of impinging
signals, it is inevitable to cause performance loss by employing
the existing channel estimation schemes \cite{Ma_J_SBL_Time_varing} due to the
power leakage caused by spatial sample mismatching.
Finally, {in \cite{Ma_J_SBL_Time_varing}}, we did not incorporate the noise covariance into the model parameters.}

%it is important that
%the reconstruction of a DL virtual channel model parameter,
%which relies on channel spatial covariance matrix,
%is just an empirical inference that can't be confirmed correctly in theory.

Different from the aforementioned works, this paper focuses on the DL channel restoration for the time-varying massive MIMO networks,
where both the TDD and FDD modes are considered.
In order to exploit the low-rank property of the spatially correlated massive MIMO channel,
we will directly learn the information of the UL channel model
instead of requiring and analyzing the channel covariance matrices.
First, an time-varying off-grid massive MIMO channel model with the adoption of Taylor series and the virtual channel representation (VCR) \cite{VCR} is constructed,
Then, a novel sparse Bayesian learning (SBL) framework is designed to estimate
the spatial signatures, the off-grid bias and temporal varying characteristics of the sparse virtual channel model as well as the observation noise covariance.
To avoid the unacceptable complexity, we apply a coordinate-wise maximization based expectation maximization (EM) algorithm to capture the parameters listed above.
Next, according to the spatial signatures,
we use a unified low dimensional Kalman filter (KF) for the virtual channel tracking.
Thanks to the reciprocity of the UL DOAs and the DL DODs for the scattering rays,
the DL spatial signatures, the time-correlation factors and the off-grid bias
can be directly obtained from the UL one.
%Meanwhile, since the DL temporal varying characteristics are closely approximates to the UL ones,
%we can treat the UL ones as the initial condition of the DL ones in the reconstruction step of DL channel model.
But unfortunately, %{although the spatial signatures, the time-correlation
%factors and the off-grid bias of DL channel model can be reconstructed with the knowledge of
%UL ones,
the other two kinds of model parameters can not be perfectly inferred from the UL ones,
as they are closely related with the carrier frequency.
Although we can still use the method used in the UL learning to capture the DL model parameters,
this would inevitablely cause a tremendous scale of overheads.
In order to avoid this obstacle,
we {resort to} the optimal Bayesian Kalman filter (OBKF) method
to accurately track the DL channel with the partially prior knowledge.
We first show the recursive equations of the DL virtual channel restoration.
Then, we employ an MCMC method to approximate some posterior effective statistics.
Finally, to obtain the posterior distribution of the noise second-order statistics, i.e., the covariance matrix, a factor graph based {sum-product algorithm} is introduced.

The rest of this paper is organized as follows. Section II gives the system model and the description of virtual channel model.
The main ideas of the Coordinate-wise Maximization based EM algorithm for model parameters learning and a concise depiction of UL channel tracking is illustrated in Section III.
Section IV presented the DL model parameters recovering and the DL virtual channel tracking by factor graph based OBKF method. The simulation results are given in Section V, and Section VI shows the conclusions.

Notations: We use lowercase (uppercase) boldface to denote vector (matrix).
$(\cdot )^T $ and $(\cdot )^H $ represent the transpose and the Hermitian transpose, respectively. $\mathbf I_N $ represents a $N \times N $ identity matrix.
$\delta (\cdot) $ is the Dirac delta function.
$\mathbb E \{\cdot \} $ is the expectation operator.
Denote $\text{tr} \{\cdot \}$ and $|\cdot | $ as the trace and the determinant of a matrix, respectively.
We use $[\mathbf A]_{i,j} $ and $\mathbf A_{:, \mathcal Q} $ (or $\mathbf A_{\mathcal Q,:}$) to represent the $(i,j) $-th entry of $\mathbf A $ and the submatrix of $\mathbf A $ which contains the columns (or rows) with the index set $\mathcal Q $, respectively.
$\mathbf x_{\mathcal Q} $ is the subvector of $\mathbf x $ formed by the entries with the index set $\mathcal Q $.
$\mathbf v \sim \mathcal{CN} (\mathbf 0, \mathbf I_N)$ means that $\mathbf v $ satisfies the complex circularly-symmetric Gaussian distribution with zero mean and covariance $\mathbf I_N $.
$\lfloor p \rfloor $ denotes the largest integer no more than $p $.
${\boldsymbol\Xi}^{(l-1)} \setminus {\boldsymbol \alpha}^{(l-1)}$ denotes the set ${\boldsymbol\Xi}^{(l-1)}$ expect the element ${\boldsymbol \alpha}^{(l-1)}$
The real component of $x $ is expressed as $\Re (x)$.
$\text{diag} (\mathbf x)$ is a diagonal matrix whose diagonal elements are formed the elements of $\mathbf x $,
while $\text{blkdiag} (\mathbf X_1, \mathbf X_2, \dots)$ is a block diagonal matrix formed by $\mathbf X_1, \mathbf X_2, \dots $.

  %and the channel estimation algorithm  does not require any knowledge of channel covariance.

\section{ System Model and Channel  Characteristics }

{We} consider an uplink multiuser massive MIMO system, where the BS is equipped with
{$N_t\gg 1$} antennas in the form of uniform linear array (ULA), and $K$ single-antenna users are randomly distributed in its coverage area.
We adopt a geometric channel model with $L$ scatters around the $k$-th user, and each scatter is supposed to dedicate a single propagation path.
Denote $\theta_{k,l,m}$ as a DOA of $k$-th user, $l$-th path and $m$-th time block,
and the BS antenna array spatial steering vector can be defined as:
\begin{equation}\label{steer_vec}
\mathbf a(\theta_{k,l,m})=\Big[1,e^{\jmath \frac{2\pi d}{\lambda}\sin(\theta_{k,l,m})},\ldots,e^{\jmath (N_t-1)\frac{2\pi d}{\lambda}\sin(\theta_{k,l,m})}\Big]^T,
\end{equation}
where $d\leq \lambda / 2$ is antenna spacing of the BS; $\lambda$  is the signal carrier wavelength.

It is assumed that the direction of arrival (DOA) of each path is quasi-static during  a block of $L_c$ channel  uses
and changes from block to block. The system sampling rate is $\frac{1}{T_s}$.
Then, the uplink channel $\mathbf h_{k,m} \in \mathbb C^{M \times 1}$ from user $k$ to the BS during the $m$-th block can be expressed as \cite{xqigao,Fleury,Raghavan}
\begin{align}
  \mathbf h_{k,m} =
%  \int_{-\infty}^{+\infty} \int_{\theta_{k^{\text{min}}, m}}^{\theta_{k^{\text{max}}}, m} \mathbf a (\theta_{k,m}) e^{\jmath 2\pi \nu mL_cT_s}
%    \hbar_{k}(\theta_{k,m},\nu) d\theta_{k,m} d\nu \notag \\
%&\approx
\int_{-\infty}^{+\infty} \sum_{l=1}^{L} \mathbf a (\theta_{k,l,m}) e^{\jmath 2\pi \nu mL_cT_s}
    \hbar_{k}(\theta_{k,l,m},\nu) d\nu,\label{eq:h_k}
\end{align}
{where} $\hbar_{k}(\theta_{k,l,m},\nu)$ is the joint
angle-Doppler channel gain function of the user $k$ corresponding to the direction of arrival (DOA) $\theta_{k,l,m}$ and the Doppler frequency $\nu$.
The channels from the BS to different users are assumed to be statistically independent.

As in \cite{JIN_MMWAVE}, the VCR can be utilized to dig the sparsity of $\mathbf h_{k,m} $ as
\begin{align}
\mathbf{\tilde h}_{k,m} = \mathbf F_{N_t} \mathbf h_{k,m},\label{eq:DFT_virtual}
\end{align}
where $\mathbf{\tilde h}_{k,m} $ is the virtual channel of $\mathbf h_{k,m} $, and $\mathbf F_{N_t}$ is the $N_t\times N_t$ normalized discrete Fourier transformation (DFT)  matrix with $(p,q)$th
entry as $[\mathbf F_{N_t}]_{p,q}=\frac{1}{\sqrt N_t} e^{-\jmath\frac{2\pi pq}{N_t}}$. Furthermore, we can adopt the simultaneously
sparse signal model to depict the dynamics of $\mathbf{\tilde h}_{k,m} $ by adopting the first order auto regressive (AR) model \cite{AR} as
\begin{align}
\left\{
\begin{aligned}
\mathbf{\tilde h}_{k,m}=&\text{diag}(\mathbf c_k)\mathbf{ r}_{k,m},\\
\mathbf{ r}_{k,m}=&\alpha_k \mathbf{ r}_{k,m-1}+ \boldsymbol\upsilon_{k,m},
\end{aligned}
\right.\label{eq:state}
\end{align}
where the time-varying processes $\mathbf{ r}_{k,m} $ represents Gaussian Markov random processes,
{$\alpha_{k}$} is the transmission factor,
$\boldsymbol\upsilon_{k,m}\sim\mathcal{CN}(0,\boldsymbol \Lambda_{k})$ is the process noise vector
where $\boldsymbol \Lambda_{k} = \text{diag}([\lambda_{k,1}^2,\lambda_{k,2}^2,\cdots, \lambda_{k,N_t}^2])$, and the spatial signature\cite{signature} vector $\mathbf c_k$ is determined by the set
 {\begin{align}
&\mathcal Q_k\!=\!\left\{p\Big|\left\lfloor N_t\frac{d}{\lambda}\sin(\theta_k^{\min})\right\rfloor \le \!p  \! \le\left\lfloor N_t\frac{d}{\lambda}\sin(\theta_k^{\max})\right\rfloor, p\in \mathbb Z \right\},  \label{eq:nonzero_Q}
\end{align}}
as $[\mathbf c_k]_p = 1 $ when $p\in \mathcal Q_k $.
%\begin{align} \label{c k}
%[\mathbf c_k]_p=
%\left\{
%\begin{aligned}
%1~~~& \text{if } p\in \mathcal Q_k,\\
%0~~~& \text{otherwise.}
%\end{aligned}
%\right.
%\end{align}

It can be checked that the locations of the non-zero elements of $\mathbf c_k$
depends on the angle spread (AS) information of the user $k$, i.e., $[\theta_k^{\min}, \theta_k^{\max}] $.
Theoretically, the AS information does not change drastically within thousands of the channel coherence time $L_t T_s $,
which means that $Q_k $ will remain time-invariant within a much longer period.
Furthermore, under the massive MIMO scenario, especially at the millimeterwave and Tera Hertz band,
the AS will be limited in one narrow region, and the number of the non-zero elements in $\mathbf c_k$, i.e., $| Q_k|$, will be much less than {$N_t$.}
Thus, the virtual channel $\mathbf{\tilde h}_{k,m}$ can be treated as suitably sparse signal.

Then we take the off-grid model into consideration.
In fact, the DFT basis {in (\ref{eq:DFT_virtual})} conducts a discrete spatial sample for the impinging
signals with general fixed sampling grid, {and} discretely covers the entire spatial angle domain.
{However}, in {the} real transition processes, the DOAs would not exactly impinging on those grids,
and the direction mismatching happens.
Under such circumstance,
define the bias vector $\boldsymbol \rho_k$,
we introduce a bias-added DFT matrix,
whose special index will be added with $\boldsymbol \rho_k$,
i.e. $p^* = p + [\boldsymbol \rho_k]_p$.
{Correspondingly,  the set} $\mathcal Q_k$  can be redefined as:
\begin{align}
&\mathcal Q_k\!=\!\left\{p\Big|p + \rho_{p} =  N_t\frac{d}{\lambda}\sin(\theta_{k,l,m}), p\in \mathbb Z \right\}, \rho_{k,l} \in [-0.5,0.5] \label{eq:true nonzero_Q}
\end{align}

\begin{figure}[!t]
	\centering
	\includegraphics[width=85mm]{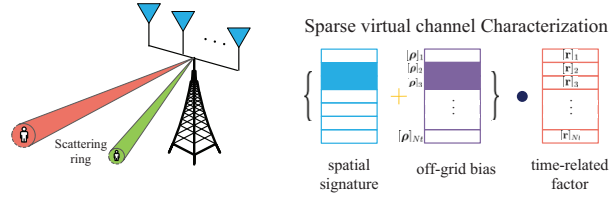}
	\caption{The system scene of our model.}
	\label{fig:system_scene2}
\end{figure}

A simple example is illustrated in \figurename{ \ref{fig:system_scene2}}.
It intuitively explains the relationship between the spatial parameters and the virtual channel vector.
%\remark{Now, we can obtain some interesting results about the virtual channel  $\mathbf{\tilde h}_{k,m}$ under the condition ${N_t\rightarrow+\infty}$: (i) The entries of $\mathbf{\tilde h}_{k,m}$ with index $p\notin\mathcal Q_k$ statistically  {tend} to be zero; (ii) The nonzero entries of $\mathbf{\tilde h}_{k,i}$ with index $p\in\mathcal Q$ are statically uncorrelated with each other. To intuitively explain the sparsity of the virtual channel vector, a simple example is illustrated in \figurename{ \ref{fig:system}}}.
%In the massive MIMO systems, the large number of antennas and the terminals will induce
%huge overhead in the downlink training or the pilot contamination in the uplink training\cite{gao_access}.
%To alleviate such general headaches, reducing the effective channel dimension is often necessary.

{Before proceeding,} we use $\mathbf A$ to represent $\mathbf F_{N_t}$ for simplicity,
{Inspired by the above observation,} the channel vector $\mathbf h_{k,m}$ can be approximated with the Taylor series expansion as
%写出rou的表达式，泰勒展开的时候它应该等于什么
%可以直接对dft矩阵乘偏移？
\begin{align}\label{vcr_h}
\mathbf h_{k,m} =[\mathbf A^H + \mathbf B^H \text{diag}(\boldsymbol \rho_k)]_{:,\mathcal Q_k} [\mathbf{\tilde h}_{k,m}]_{\mathcal Q_k}
= [\mathbf \Phi(\boldsymbol \rho_k)]_{:,\mathcal Q_k}^H [\mathbf{\tilde h}_{k,m}]_{\mathcal Q_k},
\end{align}
where $[\mathbf B^H]_{:, p}$ is obtained through taking derivative of $[\mathbf A^H]_{:, p}$
{with respect to $p$;}
{every element of $\boldsymbol \rho_k$ is the bias added on the corresponding predefined grid.}

We can rewrite the AR model of {the practical channel} as
\begin{align}
\left\{
\begin{aligned}
  \mathbf h_{k,m} &= \mathbf \Phi(\boldsymbol \rho_k)^H\text{diag}(\mathbf c_k) \mathbf{r}_{k,m}, \\
  \mathbf r_{k,m} &= \alpha_k \mathbf r_{k,m-1} + \boldsymbol \upsilon_{k,m},
\end{aligned}
\right.\label{h AR}
\end{align}
where the definitions of $\mathbf r_{k,m}$, $\alpha_k$, and
$\boldsymbol \upsilon_{k,m}$ are same with as in (\ref{eq:state}).

{Notice that, in (\ref{h AR}), $\mathbf c_k$ and $\boldsymbol \rho_k$
characterizes the spatial signatures and the AOA bias of the user $k$,
while both $\mathbf \Lambda_k$ and $\alpha_k$ depict the temporal varying characteristics of the virtual channel.
After the construction of the AR model,
the learning of the channel statistical characteristics
is equivalent to capturing the model parameters $\boldsymbol \rho_k $, $\alpha_k $, $\mathbf c_k $, $\mathbf \Lambda_k$.
Moreover, the
characteristics of the {AR} model for one specific user changes so slowly that $\mathbf \Xi_k$ is constant during a large number of the consecutive channel coherence blocks.}

\section{Model Parameters Capturing VIA Uplink Training and Uplink Channel Tracking}

Without loss of generality, we assume that the current cell is allocated with $\tau\le K$ orthogonal training sequences of length $L_s\le L_c$.
Denote the orthogonal training set as $\mathbf S=[\mathbf s_1,\mathbf s_2,\ldots,\mathbf s_{\tau}]$ with $\mathbf s_i^H\mathbf s_j=L_s\sigma_p^2\delta(i-j)$, where $\sigma_p^2$ is the pilot power.
For the ease of illustration, we assume that $ K = C\tau$, where $C $ is an integer no less than $1$.

Following most standards, there exists one long UL training period called preamble at the very beginning of each transmission.
We can use the preamble to obtain the model parameters.
Since we do not assume any prior spatial information,
we have to divide $K$ users into $C$ groups, each containing $\tau$ users such that $\tau$ orthogonal training sequences are sufficient for each group.

For the illustration simplicity, we take the first group as an example, and use $M$ channel blocks to learn the channel model parameters $\boldsymbol\Xi_k$.
The received training signal during the $m$-th block can be written as
\begin{equation}\label{recv signal mtx}
  \mathbf Y_m = \sum_{k = 1}^{\tau} {\mathbf h_{k,m} \mathbf s_k^T} + \mathbf N_m = \sum_{k = 1}^{\tau} \mathbf \Phi(\boldsymbol \rho_k)^H \text{diag}(\mathbf c_k) \mathbf r_{k,m} \mathbf s_k^T +\mathbf N_m,
\end{equation}
where $\mathbf N_m$ denotes the independent additive white Gaussian noise matrix with elements distributed as i.i.d. $\mathcal {CN}(0, \sigma_n^2)$; $\sigma_n^2$ is assumed to be unknown.
Moreover, we define the $N_t L_s \times 1$ vector $\mathbf y_m = \text{vec}(\mathbf Y_m)$ and $N_t L_s \times 1 $ vector $\mathbf n_m = \text{vec} (\mathbf N_m)$. Then (\ref{recv signal mtx}) can be rearranged as
\begin{equation}\label{recv signal vec}
  \mathbf y_m = \sum_{k = 1}^{\tau} \underbrace{(\mathbf s_k \otimes \mathbf \Phi(\boldsymbol \rho_k)^H)\text{diag} (\mathbf c_k)}_{\mathbf J_k} \mathbf r_{k,m} + \mathbf n_m = \mathbf J \mathbf r_m + \mathbf n_m,
\end{equation}
where $\mathbf n_m \sim \mathcal{CN}(\mathbf 0, \sigma_n^2 \mathbf I_{N_t L_s})$, $\mathbf J = [\mathbf J_1, \mathbf J_2, \ldots, \mathbf J_\tau] \in \mathbb C^{N_t L_s \times N_\tau}$, $\mathbf r_m = [\mathbf r_{1,m}^T, \mathbf r_{2,m}^T, \ldots, \mathbf r_{\tau,m}^T] \in \mathbb C^{N \tau \times 1}$.
For further use, we define the $N_t L_s M \times 1$ vector $\mathbf y = [\mathbf y_1^T, \mathbf y_2^T, \ldots, \mathbf y_M^T]$, the $N \tau M \times 1$ vector $\mathbf r = [\mathbf r_1^T, \mathbf r_2^T, \ldots, \mathbf r_{\tau}^T]$, the $N \tau \times 1$ vector
{$\boldsymbol \rho = [\boldsymbol \rho_1^T, \boldsymbol \rho_2^T, \ldots, \boldsymbol \rho_{\tau}^T]$,
$\tau\times 1$ vector
$\boldsymbol \alpha=[\alpha_1,\alpha_2,\ldots,\alpha_{\tau}]^T$, the $N \times 1$ vector $\mathbf c =  [\mathbf c_1^T, \mathbf c_2^T, \ldots, \mathbf c_{\tau}^T]$, and the
$\tau N\times \tau N$ matrix
$\boldsymbol\Lambda=\text{blkdiag}\{\boldsymbol\Lambda_1,\boldsymbol\Lambda_2,\ldots,\boldsymbol\Lambda_\tau\}$.}

{Here, the task of the preamble is to capture the parameter set  $\mathbf \Xi= \{ \boldsymbol \rho, \mathbf c, \boldsymbol \alpha, \boldsymbol \Lambda, \sigma_n^2 \}$ with the observation model (\ref{recv signal vec})
and the state equation (\ref{h AR}).}

\subsection{Problem Formulation}

The objective of the learning is to estimate the best fitting parameters set $\boldsymbol\Xi$ with the given observation vector $\mathbf y$.
Theoretically, the ML estimator for $\boldsymbol\Xi$ can be formulated as
\begin{align}
\boldsymbol{\hat\Xi}=&\arg \max_{1\geq \alpha_k \geq 0,~\lambda_{k,p}\geq 0,~|[\boldsymbol \rho_k]_p| < 0.5,~[\mathbf c_k]_p\in\{0,1\} } \ln p(\mathbf y;\boldsymbol\Xi).
\end{align}

Obviously, such estimator involves all possible combinations of the $\mathbf r$ and is not feasible to directly achieve the ML solution due to its high dimensional search.
Nonetheless, one alternative method is to search the solution iteratively via the EM algorithm \cite{EM_BASE}.
Furthermore, in order to achieve the faster convergence with a correspondingly lower complexity,
we will adopt the Gauss-Seidel scheme and perform the coordinate-wise maximization based EM algorithm in the following.

\subsection{Coordinate-wise Maximization based \text{EM} to Accomplish Simultaneously Sparse Signal Learning}

Similar to the classical EM algorithm, the coordinate-wise maximization based EM algorithm iteratively
produces a sequence of ${\boldsymbol\Xi}^{(l)}, l = 1,2,\ldots $, and each iteration is divided into two steps:

{\textbf{$\bullet$ Expectation step (E-step)}}
\begin{align}
Q\left({\boldsymbol \alpha},\hat{\boldsymbol\Xi}^{(l-1)}\right) &= \mathbb E_{\mathbf{ r}|\mathbf{ y}; \hat{\boldsymbol\Xi}^{(l-1)}} \big\{ \ln p\left(\mathbf{ y}, \mathbf{ r}; {\boldsymbol \alpha}, {\boldsymbol\Xi}^{(l-1)} \setminus {\boldsymbol \alpha}^{(l-1)} \right)\big\}.  \label{E alpha} \\
Q\left({\mathbf \Lambda},\hat{\boldsymbol\Xi}^{(l-1)}\right) &= \mathbb E_{\mathbf{ r}|\mathbf{ y}; \hat{\boldsymbol\Xi}^{(l-1)}} \big\{ \ln p\left(\mathbf{ y}, \mathbf{ r}; {\mathbf \Lambda}, {\boldsymbol\Xi}^{(l-1)} \setminus {\mathbf \Lambda}^{(l-1)} \right)\big\}. \label{E Lambda}\\
Q\left({\mathbf c},\hat{\boldsymbol\Xi}^{(l-1)}\right) &= \mathbb E_{\mathbf{ r}|\mathbf{ y}; \hat{\boldsymbol\Xi}^{(l-1)}} \big\{ \ln p\left(\mathbf{ y}, \mathbf{ r}; {\mathbf c}, {\boldsymbol\Xi}^{(l-1)} \setminus {\mathbf c}^{(l-1)} \right)\big\}. \label{E c}\\
Q\left({\boldsymbol \rho},\hat{\boldsymbol\Xi}^{(l-1)}\right) &= \mathbb E_{\mathbf{ r}|\mathbf{ y}; \hat{\boldsymbol\Xi}^{(l-1)}} \big\{ \ln p\left(\mathbf{ y}, \mathbf{ r}; {\boldsymbol \rho}, {\boldsymbol\Xi}^{(l-1)} \setminus {\boldsymbol \rho}^{(l-1)} \right)\big\}. \label{E rho}\\
Q\left({\sigma_n^2},\hat{\boldsymbol\Xi}^{(l-1)}\right) &= \mathbb E_{\mathbf{ r}|\mathbf{ y}; \hat{\boldsymbol\Xi}^{(l-1)}} \big\{ \ln p\left(\mathbf{ y}, \mathbf{ r}; {\sigma_n^2}, {\boldsymbol\Xi}^{(l-1)} \setminus {\sigma_n^2}^{(l-1)} \right)\big\}. \label{E sigma}
\end{align}

{\textbf{$\bullet$ Maximization step (M-step)}}
\begin{align}
\boldsymbol{\hat \alpha}^{(l)}=\arg\max_{{\boldsymbol \alpha}} Q\big({\boldsymbol \alpha},\hat{\boldsymbol\Xi}^{(l-1)}\big). \\
\mathbf{\hat \Lambda}^{(l)}=\arg\max_{{\mathbf \Lambda}} Q\big({\mathbf \Lambda},\hat{\boldsymbol\Xi}^{(l-1)}\big). \\
\mathbf{\hat c}^{(l)}=\arg\max_{{\mathbf c}} Q\big({\mathbf c},\hat{\boldsymbol\Xi}^{(l-1)}\big). \\
\boldsymbol{\hat \rho}^{(l)}=\arg\max_{{\boldsymbol \rho}} Q\big({\boldsymbol \rho},\hat{\boldsymbol\Xi}^{(l-1)}\big). \\
{\hat {\sigma_n^2}}^{(l)}=\arg\max_{{\boldsymbol \rho}} Q\big(\sigma_n^2,\hat{\boldsymbol\Xi}^{(l-1)}\big).
\end{align}

In the $l$-th iteration, the E-step is to derive those objective functions
as the expectation of the probability density function (PDF) $p(\mathbf{ y}, \mathbf{ r}; \hat{\boldsymbol\Xi} )$ over the hidden variable $\mathbf{r}$ by setting $\boldsymbol\Xi$ as the estimated model parameters $\hat{\boldsymbol\Xi}^{(l-1)}$ in the previous iteration;
the M-step is to find the new estimation ${\boldsymbol\Xi^{(l)}}$ by maximizing them.
It has been proved that the sequence $\{\boldsymbol{\hat\Xi}^{(l)}\}$ converges to a stationary point of the likelihood function \cite{EM}.

\subsection{ {Expectation step}}

In this subsection, we will carefully derive the three objective functions in \eqref{E alpha}-\eqref{E rho}.
Now, we first examine $Q\left({\boldsymbol \alpha},\hat{\boldsymbol\Xi}^{(l-1)}\right) $.
Since the received samples $\mathbf y$ are known, the objective function
$Q\left({\boldsymbol \alpha},\hat{\boldsymbol\Xi}^{(l-1)}\right) $ {can be expressed} as
\begin{align}\label{eq:Q alpha function}
  Q\big(\!{\boldsymbol \alpha},\!\hat{\boldsymbol\Xi}^{(l-1)}\!\big)
 \!\!=\! \mathbb E_{\mathbf{ r}|\mathbf{ y}; \hat{\boldsymbol\Xi}^{(l-1)}} \!\big\{\! \ln p(\mathbf{y}|\mathbf{ r}; {\boldsymbol \alpha}, {\boldsymbol\Xi}^{(l-1)} \!\setminus\! {\boldsymbol \alpha}^{(l-1)}\!) \!\big\}\!\!+\!\!\mathbb E_{\mathbf{ r}|\mathbf{ y}; \hat{\boldsymbol\Xi}^{(l-1)}} \!\big\{\! \ln p(\mathbf{ r};\! {\boldsymbol \alpha},\! {\boldsymbol\Xi}^{(l-1)} \!\setminus\! {\boldsymbol \alpha}^{(l-1)}\!) \!\big\}.
\end{align}
%\begin{align}\label{eq:Q_function}
% Q\big({\boldsymbol\Xi},\hat{\boldsymbol\Xi}^{(l-1)}\big)&=\mathbb E_{\mathbf{ r}|\mathbf{ y}; \hat{\boldsymbol\Xi}^{(l-1)}}
%\big\{ \ln p\left(\mathbf{ y}, \mathbf{ r}; {\boldsymbol\Xi} \right)\big\}\notag\\
%&=\int_{\mathbf{ r}}
%p(\mathbf{ r}|\mathbf y,\boldsymbol{\hat\Xi}^{(l-1)})\left[
%\ln p(\mathbf y|\mathbf{ r},\boldsymbol{\Xi})
%+\ln p(\mathbf{ r}|\boldsymbol{\Xi})\right]d{\mathbf{ r}}\notag\\
%&=\mathbb E_{\mathbf{ r}|\mathbf{ y}; \hat{\boldsymbol\Xi}^{(l-1)}}\left\{\ln p(\mathbf{y}|\mathbf{ r};\boldsymbol{\Xi}) \right\} +
%\mathbb E_{\mathbf{ r}|\mathbf{ y}; \hat{\boldsymbol\Xi}^{(l-1)}}\left\{\ln p(\mathbf{ r};\boldsymbol{ \Xi}) \right\}.
%\end{align}

From  (\ref{recv signal vec}), we get the conditional PDF as:
\begin{align}\label{eq:y con r}
  p(\mathbf{y}_m|\mathbf{ r}_m; {\boldsymbol \alpha}, {\boldsymbol\Xi}^{(l-1)} \setminus {\boldsymbol \alpha}^{(l-1)}) \sim \mathcal{CN} \left(\sum_{k=1}^\tau(\mathbf J_k \mathbf{ r}_{k,m}\big), \sigma_n^2 \mathbf I_{N_t L_s}\right).
\end{align}

Meanwhile, we have
\begin{align}\label{eq:joint_r_PDF}
\ln p(\mathbf{ r}; {\boldsymbol \alpha}, {\boldsymbol\Xi}^{(l-1)} \setminus {\boldsymbol \alpha}^{(l-1)})
&= \sum_{k = 1}^{\tau} {\ln{p\Big(\mathbf{ r}_{k,1}\Big)}} + \sum_{m = 2}^{M} \sum_{k = 1}^{\tau} {\ln {p\Big(\mathbf{ r}_{k,m}|\mathbf{ r}_{k,m-1};
    {\boldsymbol \alpha}, {\boldsymbol\Xi}^{(l-1)} \setminus {\boldsymbol \alpha}^{(l-1)}\Big)}},
\end{align}
where the conditional PDF $p\Big(\mathbf{ r}_{k,m}|\mathbf{ r}_{k,m-1}; {\boldsymbol \alpha}, {\boldsymbol\Xi}^{(l-1)} \!\! \setminus \!\! {\boldsymbol \alpha}^{(l-1)}\Big)$
can be written as
\begin{align}\label{eq:cond_r_PDF}
p\Big(\mathbf{ r}_{k,m}|\mathbf{ r}_{k,m-1}; {\boldsymbol \alpha}, {\boldsymbol\Xi}^{(l-1)} \!\!\setminus \!\! {\boldsymbol \alpha}^{(l-1)}\Big)
=\frac{\exp \left( - (\mathbf{ r}_{k,m}\!-\!\alpha_{k}\mathbf{ r}_{k,m-1})^H
(\boldsymbol {\hat \Lambda}_k^{(l-1)})^{-1}\!(\mathbf{ r}_{k,m}\!-\!\alpha_{k}\mathbf{ r}_{k,m-1}) \right)}{{\pi}^N |\boldsymbol {\hat \Lambda}_k^{(l-1)}| }.
\end{align}

{Before proceeding,
we define three posterior statistics about $\mathbf{{ r} }_{m}$, i.e.,
$\mathbf{\hat{ r} }_{k,m}^{(l-1)} \!\stackrel{\vartriangle}{=}\! \mathbb E\Big\{\mathbf{ r}_{k,m}|\mathbf y,\boldsymbol{\hat\Xi}^{(l-1)}\Big\}$,
$\boldsymbol\Theta_{k,m}^{(l-1)}\stackrel{\vartriangle}{=}  \mathbb E\Big\{\mathbf{ r}_{k,m}\mathbf{ r}_{k,m}^H|\mathbf y,\boldsymbol{\hat\Xi}^{(l-1)}\Big\}$, and
$\boldsymbol\Pi_{k,m-1,m}^{(l-1)}\stackrel{\vartriangle}{=}\mathbb E\Big\{\mathbf{ r}_{k,m-1}\mathbf{ r}_{k,m}^H|\mathbf y,\boldsymbol{\hat\Xi}^{(l-1)}\Big\}$.
Then, plugging \eqref{eq:y con r}-\eqref{eq:cond_r_PDF} into \eqref{eq:Q alpha function} and taking some reorganizations, we can obtain:}
\begin{align}\label{eq:final Q alpha}
  Q(\alpha_k, \hat{\boldsymbol\Xi}^{(l-1)})
%=&\mathbb E\left\{ \sum_{m=2}^{M} - \ln|(\boldsymbol {\hat \Lambda}_k^{(l-1)})|-
% \left((\mathbf{ r}_{k,m}-\alpha_k \mathbf{ r}_{k,m-1})^H
%(\boldsymbol {\hat \Lambda}_k^{(l-1)})^{-1}
%(\mathbf{ r}_{k,m}-\alpha_k \mathbf{ r}_{k,m-1})\right)\right\}\notag\\
%%%%%%%%%%%%%%%%%%%%%%%%%%%%%%%%%%%%%%%%%%%%%%%%%%%%%%%%%%
=&\sum_{m=2}^{M} - \alpha_k^2\tr\Big\{ \Big({\boldsymbol {\hat \Lambda}_k^{(l-1)}}\Big)^{-1}
\boldsymbol\Theta_{k,m-1}^{(l-1)}\Big\}+ 2\alpha_k\Re\Big\{\tr\Big\{ \Big({\boldsymbol {\hat \Lambda}_k^{(l-1)}}\Big)^{-1}
\boldsymbol\Pi_{k,m-1,m}^{(l-1)}\Big\}\Big\} \!+\!C_1,
%%%%%%%%%%%%%%%%%%%%%%%%%%%%%%%%%%%
\end{align}
where $C_1$ is the sum of the items not related with $\alpha_k$.

By doing similar process of \eqref{eq:y con r}-\eqref{eq:final Q alpha}, we can derive other objective functions as follows.
\begin{small}
\begin{align}
  Q(\mathbf \Lambda_k, \hat{\boldsymbol\Xi}^{(l-1)})
=& \sum_{m=2}^{M}\Big( -\ln |\boldsymbol {\Lambda}_k| -\tr\Big(\boldsymbol {\Lambda}_k^{-1} \boldsymbol\Theta_{k,m}^{(l-1)} \Big)- \Big({\widehat\alpha_k}^{{(l-1)}}\Big)^2 \tr\Big(\boldsymbol {\Lambda}_k^{-1}
\boldsymbol\Theta_{k,m-1}^{(l-1)}\Big) \notag\\
&+ 2 {\widehat \alpha_k}^{{(l-1)}} \Re\Big\{ \tr\big(\boldsymbol {\Lambda}_k^{-1}
 \boldsymbol\Pi_{k,m-1,m}^{(l-1)}\big)\Big\}\Big) + C_2, \label{eq:final Q Lambda}\\
 %%%%%%%%%%%%%%%%%%%%%%%%%%%%%%%%%%%%%%%%%%%%%%%%%%%%%%%%%%%%%%%%%%%%%%%%%%%%%%
Q(\mathbf c_k,  \hat{\boldsymbol\Xi}^{(l-1)})=&
\frac{2}{\big({\hat \sigma_n}^{{(l-1)}}\big)^2}\Big(  {\sum\limits_{m=1}^{M}
\Re\Big\{\mathbf y_m^H[\mathbf s_k\otimes\boldsymbol \Phi^H(\boldsymbol {\hat \rho}_k^{(l-1)})]\text{diag}
\big(\widehat{\mathbf r}_{k,m}^{(l-1)}\big)\Big\}} \Big) \mathbf c_k \notag \\
&- \frac{\|\mathbf s_k\|^2}{\big({\hat \sigma_n}^{{(l-1)}}\big)^2}\mathbf c_k^T \Big(   \sum\limits_{m=1}^{M}
\Big( [ \boldsymbol \Phi^H(\boldsymbol { \hat \rho}_k^{(l-1)}) \boldsymbol \Phi(\boldsymbol {\hat \rho}_k^{(l-1)})] \odot \boldsymbol\Theta_{k,m}^{(l-1)}\odot
\mathbf I \Big) \Big) \mathbf c_k + C_3, \label{eq:final Q c}
\end{align}
\begin{align}
Q(\boldsymbol \rho_k,  \hat{\boldsymbol\Xi}^{(l-1)})
=&
\frac{2}{\big({\hat \sigma_n}^{{(l-1)}}\big)^2}\Big(\sum\limits_{m=1}^{M}\Re\left\{\mathbf y_m^H(\mathbf s_k^H \otimes\mathbf B^H)\text{diag}(\widehat{\mathbf r}_{k,m}^{(l-1)}) \text{diag}(\mathbf {\hat c}_k^{(l-1)}) \Big\} \right) \boldsymbol \rho_k   \notag \\
&-\frac{2\|\mathbf s_k\|^2}{\big({\hat \sigma_n}^{{(l-1)}}\big)^2}
\Big(\sum\limits_{m=1}^{M} (\mathbf {\hat c}_k^{(l-1)})^T
\left(\Re \Big\{ \mathbf A \mathbf B^H \odot \boldsymbol\Theta_{k,m}^{(l-1)}\Big\} \right)
\odot \mathbf I  \Big) \boldsymbol \rho_k \notag \\
& -
\frac{\|\mathbf s_k\|^2}{\big({\hat \sigma_n}^{{(l-1)}}\big)^2} \boldsymbol \rho_k^T \left( \sum\limits_{m=1}^{M}
\left( [\text{diag}(\mathbf {\hat c}_k^{(l-1)}) \mathbf B \mathbf B^H \text{diag} (\mathbf {\hat c}_k^{(l-1)}) ]
\odot \boldsymbol\Theta_{k,m}\right) \odot \mathbf I  \right) \boldsymbol \rho_k + C_4, \label{eq:final Q rho}\\
%%%%%%%%%%%%%%%%%%%%%%%%%%%%%%%%%%%%%%%%%%%%%%%%%%%%%%%%%%%%%%%%%%%%%%%%%%%%%%
Q(\sigma_n^2,  \hat{\boldsymbol\Xi}^{(l-1)})
=& \!\!-\! \frac{1}{ \sigma_n^2}\!{\sum\limits_{m=1}^{M}\sum\limits_{k=1}^\tau	\|\mathbf s_k\|^2\! \tr \left\{ \boldsymbol \Psi(\boldsymbol {\hat \rho}_k^{(l-1)}, \mathbf {\hat c}_k^{(l-1)})  \boldsymbol \Theta_{k,m}^{(l-1)}\right\} } \notag \\
 &+ \frac{2}{ \sigma_n^2} \sum\limits_{m=1}^{M}\Re\Big\{\mathbf y_m^H\sum\limits_{k=1}^\tau \mathbf{\widehat J}_k^{(l-1)}
	\mathbf{\widehat r}_{k,m}^{(l-1)} \Big\}- {N_t L_s}\sum_{m = 1}^{M} {\ln{\pi \sigma_n^2}} - \frac{1}{\sigma_n^2}\sum_{m=1}^{M}{\mathbf y_m^H \mathbf y_m} + C_5. \label{eq:final Q sigma}
\end{align}
\end{small}
where ${\mathbf {\widehat J}_k^{(l-1)}}\!=\!{(\mathbf s_k \!\otimes\! \mathbf \Phi\!(\boldsymbol {\widehat\rho}_k^{(l-1)})^H)\text{diag} (\mathbf {\widehat c}_k^{(l-1)})}$, $\boldsymbol \Psi(\boldsymbol \rho_k, \mathbf c_k) = \text{diag}(\mathbf c_k) \boldsymbol \Phi(\boldsymbol \rho_k) \boldsymbol \Phi(\boldsymbol \rho_k)^H \text{diag}(\mathbf c_k)$, and $C_2$, $C_3$, $C_4$, $C_5 $ are not related to their own objective parameter.

From \eqref{eq:final Q alpha}-\eqref{eq:final Q sigma}, it can be found that those expectation functions are dependent on
$\mathbf{\widehat r}_{k,m}^{(l-1)}$, $\boldsymbol\Theta_{k,m}^{(l-1)}$, and
$\boldsymbol\Pi_{k,m-1,m}^{(l-1)}$. Similar to \cite{Ma_J_SBL_Time_varing}, with given $\mathbf y$ and $\boldsymbol{\hat\Xi}^{(l-1)}$,
the above three terms can be achieved from the following
state-space model as
\begin{align}
\mathbf{ r}_{m}&= \mathbf{\widehat X}^{(l-1)}  \mathbf{ r}_{m-1}+   \boldsymbol\upsilon_{m},\label{eq:dss_state}\\
\mathbf y_m&=\mathbf {\widehat J}^{(l-1)} \mathbf{ r}_{m}+\mathbf n_m,\label{eq:dss_measure}
\end{align}
where $\boldsymbol\upsilon_{m}  =[\boldsymbol\upsilon_{1,m}^T,\boldsymbol\upsilon_{2,m}^T,\cdots, \boldsymbol\upsilon_{\tau,m}^T]^T \sim\mathcal{CN}(0,\widehat {\boldsymbol \Lambda}^{(l-1)})$,
$\mathbf n_m \sim \mathcal{CN}\left(0, {\hat \sigma}^{2^{(l-1)}} \otimes \mathbf I_{N_t \tau}\right) $
\begin{align}
&\mathbf{\widehat X}^{(l-1)}   = \text{diag}(\hat\alpha_1^{(l-1)} , \hat\alpha_2^{(l-1)} , \cdots, \hat\alpha_{\tau}^{(l-1)}) \otimes \mathbf I_{N_t}, \\
&\boldsymbol{\widehat \Lambda} ^{(l-1)}= \text{blkdiag}(\hat{\boldsymbol\Lambda}_1^{(l-1)},\hat{\boldsymbol\Lambda}_2^{(l-1)},\cdots,\hat{\boldsymbol\Lambda}_{\tau}^{(l-1)}),\\
&\mathbf {\widehat J} ^{(l-1)}= \text{blkdiag}(\mathbf{\widehat J}_1^{(l-1)},
\mathbf{\widehat J}_2^{(l-1)},\cdots,\mathbf{\widehat J}_{\tau}^{(l-1)}).
\end{align}

\subsection{{Maximization step}}

In this step, we will derive $\boldsymbol{\hat\Xi}^{(l)}$ through maximizing all the objective function of all users one by one.
%\begin{align}
%\hat {\alpha}_k^{(l)} &=\arg \max_{\alpha_k} \left\{  Q\big(\alpha_k, \hat{\boldsymbol\Xi}^{(l-1)}\big)  \right\},\tag{P1}\label{optimal_ alpha_}\\
%%%%%%%%%%%%%%%%%%%%%%%%%%%%%%%%%%%%%%%%%%%%%%%%%%%%%%%%%%%%
%{\hat {\mathbf \Lambda}}_k^{(l)} &= \arg \max_{\mathbf \Lambda_k } \left\{  Q\big(\mathbf \Lambda_k,\hat{\boldsymbol\Xi}^{(l-1)}\big)  \right\},\tag{P2}\label{optimal_lamda}\\
%%%%%%%%%%%%%%%%%%%%%%%%%%%%%%%%%%%%%%%%%%%%%%%%%%%%%%%%%%%
%\boldsymbol{\hat{ \mathbf c}_k }^{(l)} &= \arg\max_{{ \mathbf c_k }}\left\{ Q\big(\mathbf c_k, \hat{\boldsymbol\Xi}^{(l-1)}\big) \right\},\tag{P3}\label{optimal_c} \\
%%%%%%%%%%%%%%%%%%%%%%%%%%%%%%%%%%%%%%%%%%%%%%%%%%%%%%%%
%\hat {\boldsymbol \rho_k}^{(l)} &= \arg\max_{\boldsymbol \rho_k}\left\{  Q\big({\boldsymbol \rho_k}, \hat{\boldsymbol\Xi}^{(l-1)}\big)  \right\},\tag{P4}\label{optimal_ rho}\\
%%%%%%%%%%%%%%%%%%%%%%%%%%%%%%%%%%%%%%%%%%%%%%%%%%%%%%%%%
%{\hat \sigma_n}^{2^{(l)}} &=\arg\max_{\sigma_n^2}\left\{  Q\big( \sigma_n^2, \hat{\boldsymbol\Xi}^{(l-1)}\big)  \right\},\tag{P5}\label{optimal_ sigma^2}
%\end{align}
%
%\begin{align}
%\boldsymbol{\hat\Xi}^{(l)}=\arg\max_{{\boldsymbol\Xi}}\left\{Q\big({\boldsymbol\Xi},\hat{\boldsymbol\Xi}^{(l-1)}\big)\right\}.
%\tag{P1}
%\end{align}
As shown in \eqref{eq:final Q alpha}-\eqref{eq:final Q sigma}, {${\boldsymbol \Xi_k}$ of different users are uncoupled,} which means that the parameters for each user's dynamic virtual channel
can be studied independently  from user to user.
Therefore, we will solve the maximal problem above one by one and solve them for each users independently.

\subsubsection {\textbf{Searching $  \boldsymbol{\hat{ \mathbf c} }_k^{(l)}$}}
{It can be checked that $[\mathbf \Lambda_k]_{j,j}$ is nearly 0 when $j \notin \mathcal Q_k$.
Based on this observation, we use a wise search algorithm to find a solution for $ \boldsymbol{\hat{ \mathbf c} }_k^{(l)}$.}

\begin{figure}[htbp]
  \centering
  \subfigure[]
  {\label{search_c1}
  \begin{minipage}{60mm}
  \centering
  \includegraphics[width=60mm]{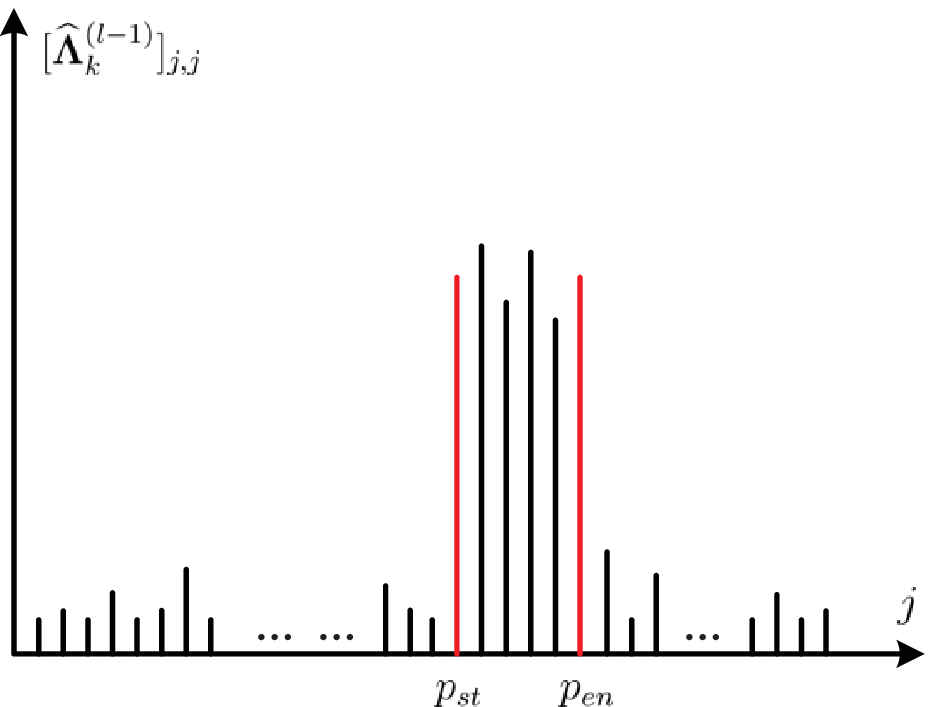}
  \end{minipage}
  }
  \subfigure[]
  {\label{search_c2}
  \begin{minipage}{60mm}
  \centering
  \includegraphics[width=60mm]{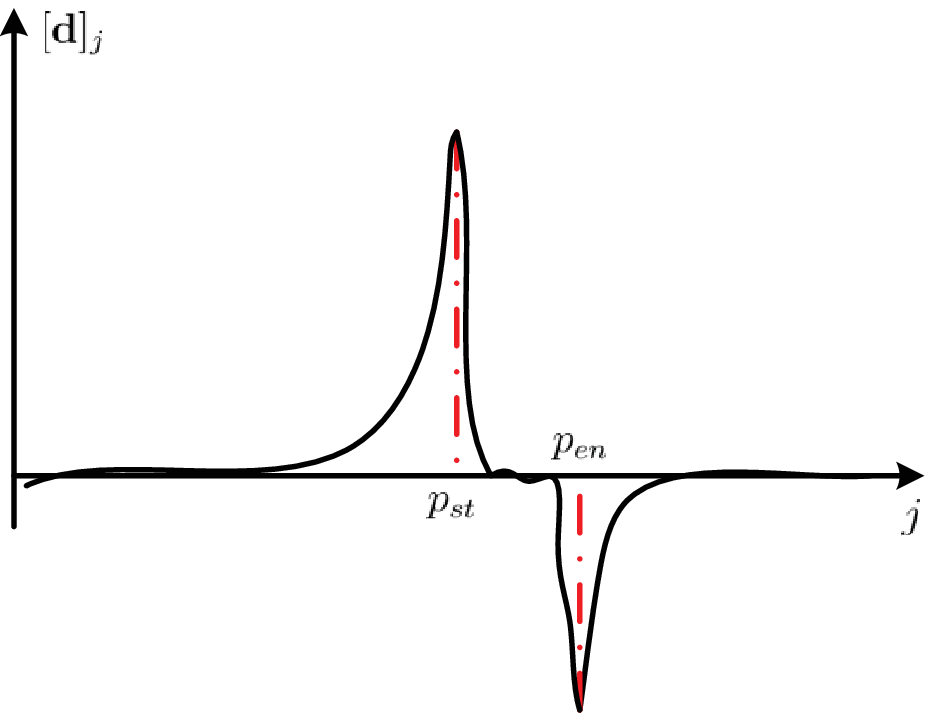}
  \end{minipage}
  }
  \caption{(a) An illustration for the position of non-zero diagonal elements in ${\widehat {\mathbf \Lambda}}_k^{(l-1)}$.
  (b) An image expression for algorithm \ref{alg:seraching_c}. }
\end{figure}

We can obtain that ${\widehat {\mathbf \Lambda}}_k^{(l-1)}$ only has a few continuous non-zero elements at its diagonal,
while its other diagonal elements are nearly zero.  \figurename {\ref{search_c1}} shows the sketch for diagonal elements of ${\widehat {\mathbf \Lambda}}_k^{(l-1)}$.
If we obtain the position of those non-zero point, we will find the optimal solution.
An easy alternative method is to obtain the position of a big increment and the position of a big decrement through forward search.
But, {since} there could be some unpredictable perturbations at those non-zero points,
the above method may cause a level of bias.

%\begin{figure}
%  \centering
%  % Requires \usepackage{graphicx}
%  \includegraphics[width=80mm]{spatial_signature1.eps}\\
%  \caption{An illustration for the position of non-zero diagonal elements in ${\widehat {\mathbf \Lambda}}_k^{(l-1)}$. }\label{search_c1}
%\end{figure}
%
%\begin{figure}
%  \centering
%  % Requires \usepackage{graphicx}
%  \includegraphics[width=80mm]{spatial_signature2.eps}\\
%  \caption{An image expression for algorithm \ref{alg:seraching_c}. }\label{search_c2}
%\end{figure}

Thus, we adopt a flattening way to avoid the influence, as shown in {algorithm \ref{alg:seraching_c}}.
First, we set all entries of $\mathbf c_k $ to zero.
Denote $s1 = [{\widehat {\mathbf \Lambda}}_k^{(l-1)}]_{j,j} + [{\widehat {\mathbf \Lambda}}_k^{(l-1)}]_{j+1,j+1} + [{\widehat {\mathbf \Lambda}}_k^{(l-1)}]_{j+2,j+2}$
and $s2 = [{\widehat {\mathbf \Lambda}}_k^{(l-1)}]_{j+3,j+3} + [{\widehat {\mathbf \Lambda}}_k^{(l-1)}]_{j+4,j+4} + [{\widehat {\mathbf \Lambda}}_k^{(l-1)}]_{j+5,j+5}$,
then we compare the two value.
Denote $[\mathbf d]_j = \ln(\frac{s2}{s1}) $ as a logarithmic function for $\frac{s2}{s1} $
and track it.
When it reaches the highest value, we set the current $j+3$ as the starting point $p_{st}$.
Continue tracking the value until it reaches its lowest value,
and set the current $j+3$ as the ending point $p_{en}$.
Then set all the elements between $[\mathbf c_k]_{p_{st}}$ and $[\mathbf c_k]_{p_{en}}$ as $1 $.
\figurename {\ref{search_c2}} shows the position searching part of algorithm \ref{alg:seraching_c}.

\begin{algorithm}
	\caption{Searching $\hat {\mathbf c}_k^{(l)}$}
	\label{alg:seraching_c}
	\renewcommand{\arraystretch}{0.9}
	\begin{algorithmic}[1]
        \STATE {\bf Input:} ${\widehat {\mathbf \Lambda}}_k^{(l-1)}$.
        \STATE {\bf Initialize:} $\hat {\mathbf c}_k^{(l)} = \mathbf 0_{N_t}$, $\max = 0$, $\min = 0$.
        \FOR {$j = 1, 2, \ldots, N_t-5 $}
        \STATE $s1 = [{\widehat {\mathbf \Lambda}}_k^{(l-1)}]_{j,j} + [{\widehat {\mathbf \Lambda}}_k^{(l-1)}]_{j+1,j+1} + [{\widehat {\mathbf \Lambda}}_k^{(l-1)}]_{j+2,j+2}$.
        \STATE $s2 = [{\widehat {\mathbf \Lambda}}_k^{(l-1)}]_{j+3,j+3} + [{\widehat {\mathbf \Lambda}}_k^{(l-1)}]_{j+4,j+4} + [{\widehat {\mathbf \Lambda}}_k^{(l-1)}]_{j+5,j+5}$.
        \STATE $[\mathbf d]_j = \ln(\frac{s2}{s1}) $.
        \ENDFOR
        \STATE $\max = [\mathbf d]_1$, $\min = [\mathbf d]_1$, $p_{st} = 1$, $p_{en} = 1$.
        \FOR {$j = 1, 2, \ldots, N_t-5 $}
        \IF {$[\mathbf d]_j < \min $}
        \STATE $\min = [\mathbf d]_j $, $p_{st} = j$.
        \ELSIF {$[\mathbf d]_j > \max $}
        \STATE $\max = [\mathbf d]_j $, $p_{en} = j$.
        \ENDIF
        \ENDFOR
        \FOR {$p = 1, 2, \ldots, N_t $}
        \IF {$p_{st} \leq p \leq p_{en}$}
        \STATE $[\hat {\mathbf c}_k^{(l)}]_p = 1 $.
        \ENDIF
        \ENDFOR
        \RETURN { $\hat {\mathbf c}_k^{(l)}$. } 	
	\end{algorithmic}
\end{algorithm}

\subsubsection{\textbf{Computing $\hat {\boldsymbol \rho_k}^{(l)}$}}
{Taking the derivatives of \eqref{eq:final Q rho} with respect to $ [\boldsymbol \rho_k]_j$, we have}
\begin{align}\label{der_rho}
  \frac{\partial Q\big(\boldsymbol \rho_k,\hat{\boldsymbol\Xi}^{(l-1)}\big) }{\partial [\boldsymbol \rho_k]_j}
  &= \frac{2}{{\hat \sigma_n}^{2^{(l-1)}}} \sum_{m=1}^{M} \Re{ \Big\{ [ \text{diag}(\mathbf {\hat c}_k^{(l-1)}) \text{diag} (\widehat {\mathbf r}_{k,m})^H (\mathbf s_k^H \otimes \mathbf B) \mathbf y_m]_j \Big\} } \notag \\
  &- \frac{2 \|\mathbf s_k\|^2}{{\hat \sigma_n}^{2^{(l-1)}}} \sum\limits_{m=1}^{M}
\Re \{ [\mathbf B^H \mathbf A \odot \boldsymbol \Theta_{k,m}^H ]_{j,j}\} [{\mathbf {\hat c}}_k^{(l-1)}]_j \notag \\
&- \frac{2 \|\mathbf s_k\|^2}{{\hat \sigma_n}^{2^{(l-1)}}} \sum\limits_{m=1}^{M}
\left[ ( \text{diag}({\mathbf {\hat c}}_k^{(l-1)}) \mathbf B^H \mathbf B \text{diag} ({\mathbf {\hat c}}_k^{(l-1)}) ) \odot \boldsymbol \Theta_{k,m} \right]_{j,j} [\boldsymbol \rho_k]_j.
\end{align}

Then, $[\hat {\boldsymbol \rho}_k^{(l)}]_j$ can be achieved by setting the  derivatives  to zero, i.e.,
 $\frac{\partial Q\big(\boldsymbol \rho_k,\hat{\boldsymbol\Xi}^{(l-1)}\big) }{\partial [\boldsymbol \rho_k]_j} = \mathbf 0$,
 and the {rough solution} $[\hat {\boldsymbol \rho}_k^{(l)}]_j^*$ can be computed as:
\begin{align}\label{update_rho}
  [\hat {\boldsymbol \rho}_k^{(l)}]_j^* =
  \frac{\sum_{m=1}^{M} \Re{ \Big\{ [ \text{diag}(\mathbf {\hat c}_k^{(l-1)}) \text{diag} (\widehat {\mathbf r}_{k,m})^H (\mathbf s_k^H \otimes \mathbf B) \mathbf y_m]_j
  \!\!-\!\!\|\mathbf s_k\|^2 [\mathbf B \mathbf A^H \odot \boldsymbol \Theta_{k,m}^H ]_{j,j} [{\mathbf {\hat c}}_k^{(l-1)}]_j  \Big\} } }
  {\|\mathbf s_k\|^2 \sum\limits_{m=1}^{M} \left[ ( \text{diag}({\mathbf {\hat c}}_k^{(l-1)}) \mathbf B \mathbf B^H \text{diag} ({\mathbf {\hat c}}_k^{(l-1)}) ) \odot \boldsymbol \Theta_{k,m} \right]_{j,j}}
\end{align}
{With} the constraint that $ [{\boldsymbol \rho}_k]_j \in [-\frac{1}{2}, \frac{1}{2}]$, so if $[\hat {\boldsymbol \rho}_k^{(l)}]_j^* \geq \frac{1}{2} $ or $[\hat {\boldsymbol \rho}_k^{(l)}]_j^* \leq -\frac{1}{2}$, the result of $\hat {\boldsymbol \rho}_k^{(l)}$ should be {bounded} as $\frac{1}{2} $ and $-\frac{1}{2} $, respectively.
%\begin{equation}
%  [\hat {\boldsymbol \rho}_k^{(l)}]_i =
%  \left\{
%  \begin{aligned}
%  &\frac{1}{2}, \qquad \qquad & [\hat {\boldsymbol \rho}_k^{(l)}]_j^* \geq \frac{1}{2} \\
%  &-\frac{1}{2}, \qquad \quad  & [\hat {\boldsymbol \rho}_k^{(l)}]_j^* \leq -\frac{1}{2} \\
%  &[\hat {\boldsymbol \rho}_k^{(l)}]_j^*, \qquad \qquad &\text{else}
%  \end{aligned}
%  \right.
%\end{equation}

\subsubsection{\textbf{Computing $\widehat {\alpha}_k^{(l)}$, ${\widehat {\mathbf \Lambda}}_k^{(l)}$,and $(\widehat \sigma_n^{(l)})^2$}}
{The computation of these three parameters is much easier than the above one.
%Take the derivatives of \eqref{eq:final Q alpha}, \eqref{eq:final Q Lambda}, and \eqref{eq:final Q sigma} with respect to $\alpha_k$, $[\boldsymbol\Lambda_k]_{j,j}$, and $\sigma_n^2$ respectively,
%then the estimation of these parameters can be achieved by setting the derivatives to zero.
After some calculations, we can obtain $\widehat {\alpha}_k^{(l)}$, ${\widehat {\mathbf \Lambda}}_k^{(l)}$,and $(\widehat \sigma_n^{{(l)}})^2$ as:}
\begin{small}\begin{align}
\hat \alpha_k^{(l)}
=&\frac{\sum\limits_{i=2}^{M} \Re\left\{\tr\left((\hat{\boldsymbol\Lambda}_k^{(l-1)})^{-1}\boldsymbol\Pi_{k,i-1,i}^{(l-1)} \right)\right\}}{\sum\limits_{i=2}^{M}  \tr\left((\hat{\boldsymbol\Lambda}_k^{(l-1)})^{-1}\boldsymbol\Theta_{k,i-1}^{(l-1)}\right)},
\label{update_alpha}\\
{[\hat{\boldsymbol\Lambda}_k]_{j,j}^{(l)}}
=&\frac{1}{M-1}\sum_{i=2}^{M} \Big[ [\boldsymbol\Theta_{k,i}^{(l-1)}]_{j,j}+(\hat{\alpha }_k^{(l-1)})^2 [\boldsymbol\Theta_{k,i-1}^{(l-1)}]_{j,j}  - 2\hat{\alpha}_k^{(l-1)} \Re\{[\boldsymbol{\Pi}_{k,i-1,i}^{(l-1)}]_{j,j}\}  \Big],
\label{update_Lambda}\\
  (\widehat \sigma_n^{{(l)}})^2 =&\frac{1}{MN_tL_s}  \Big(\sum\limits_{m=1}^{M}
  \Big(\sum\limits_{k=1}^{\tau} \|\mathbf s_k\|^2\! \tr \big\{ \boldsymbol \Psi(\boldsymbol {\hat \rho}_k^{(l-1)}, \mathbf {\hat c}_k^{(l-1)})
    \boldsymbol \Theta_{k,m}^{(l-1)} \big\}
  \!\! - 2\Re\Big\{\mathbf y_m^H\sum\limits_{k=1}^\tau \mathbf{\widehat J}_k^{(l-1)} \mathbf{\widehat{ r}}_{k,m} \Big\}
  \!\! +\mathbf y_m^H \mathbf y_m\Big)\Big). \label{update_sigma}
\end{align}\end{small}

\subsection{UL virtual channel tracking}

{Once} the parameters of the virtual channel model $\boldsymbol\Xi_k=\{\alpha_k,\boldsymbol\Lambda_k,\mathbf c_k, \boldsymbol \rho_k, \sigma_n^2\}$ have been captured in the learning phase,
%and vary on the order of seconds, while the channel coherence time $L_c T_s $ is usually in the order of millisecond\cite{gao2015spatially}, they can be deemed as unchanged variables in the long scale of coherence time,
%we can use them to track the virtual channel $\mathbf r_{k,m} $ block by block.
 the users can  be divided into different groups according to their spatial signatures
 { to remove the pilot contamination}  and realize the simultaneous training of {different} users with less orthogonal training sequences.
Specifically, the users are allocated to the same group if their spatial signatures do not overlap i.e.,
\begin{align}\label{grouping_UL}
\mathbf c_k \mathbf c_j^T =0.
\end{align}

Assume that all users are divided into $G$ groups according to \eqref{grouping_UL} and collect user indexes in the $g$-th group into the set $\mathcal G_g$.
Since the users in the same group are separated by different spatial signatures, we can assign the same training sequences for the users in one group to estimate the virtual channel $\tilde{\mathbf h}_{k,m}$.
However, different user groups will utilize orthogonal training sequences.
Therefore, we can construct a $G\times G$ matrix $\mathbf S_G$ with $\mathbf S_G^H \mathbf S_G = G\sigma_p^2\mathbf I_G$.
Then, $\mathbf s_g = [\mathbf S_G ]_{:,g}$  will be allocated to the group $g$, and all $K$ users send their training sequences simultaneously.
Thus, the received signals at the BS can be expressed as
\begin{align}
\mathbf Y_m = \sum_{g=1}^{G} \sum_{k\in \mathcal G_g} \mathbf h_{k,m} \mathbf s_{g}^H + \mathbf N_m.
\end{align}

Notice that each user in the same group have different spatial signatures.
%Denote $\mathcal Q = \{\mathcal Q_1, \mathcal Q_2, \ldots\}$.
Since $\mathbf s_g$  {is orthogonal to $\mathbf s_{g^\prime}$, $g\neq g^\prime$}, the signals for the group $g$ can be extracted as
\begin{align}
\mathbf y_{g,m}
&= \frac{1}{G \sigma_p^2} \mathbf Y_m \mathbf s_g
= \sum_{k\in \mathcal G_g} \mathbf h_{k,m} + \frac{1}{G \sigma_p^2} \mathbf N_m \mathbf s_g \notag \\
&= \sum_{k\in \mathcal G_g} [\mathbf \Phi(\boldsymbol \rho_k)]_{:, \mathcal Q_k} [\mathbf r_{k,m}]_{\mathcal Q_k} + \frac{1}{G \sigma_p^2} \mathbf N_m \mathbf s_g
= \mathbf D_{\mathcal Q} \mathbf r_{m_{\mathcal Q}} + \tilde{\mathbf n}_m.
\end{align}
where $\mathbf D_{\mathcal Q} = \left[ [\mathbf \Phi(\boldsymbol \rho_1)]_{:, \mathcal Q_1}, [\mathbf \Phi(\boldsymbol \rho_2)]_{:, \mathcal Q_2}, \ldots \right]$,
$ \mathbf r_{m_{\mathcal Q}} = \left[[\mathbf r_{1,m}]_{\mathcal Q_1}^H, [\mathbf r_{2,m}]_{\mathcal Q_2}^H, \ldots \right]^H$,
and $\tilde{\mathbf n}_m = \frac{1}{G \sigma_p^2} \mathbf N_m \mathbf s_g$ is the equivalent Gaussian white noise vector.

Define
$\boldsymbol \alpha^{*} = \text{blkdiag} \{ \text{diag}(\underbrace{\alpha_1, \alpha_1, \ldots}_{\mathcal Q_1 }), \text{diag}(\underbrace{\alpha_2, \alpha_2, \ldots}_{\mathcal Q_2 }), \ldots \}$,
$\mathbf \Lambda^{*} = \text{blkdiag} \{\mathbf [\Lambda_1]_{\mathcal Q_1, \mathcal Q_1}, [\Lambda_2]_{\mathcal Q_2, \mathcal Q_2}, \ldots \}$,
and then we can derive the following state-space model with reduced dimension to build to track $\mathbf r_{m_{\mathcal Q_k}}$, by which we can obtain the estimation of $\mathbf h_{k,m}$.
\begin{align}
%[\mathbf r_{k,m}]_{\mathcal B_k} =& \alpha_{k} [\mathbf r_{k,m-1}]_{\mathcal B_k} + [{\mathbf v}_{k,m}]_{\mathcal B_k},\\
\mathbf r_{m_{\mathcal Q}} =& \boldsymbol \alpha^{*} \mathbf r_{m_{\mathcal Q}} + \mathbf v_{m_{\mathcal Q}}^{*},\\
\mathbf y_{g,m} =& \mathbf D_{\mathcal Q} \mathbf r_{m_{\mathcal Q}} + \tilde{\mathbf n}_{k,m}. \label{eq:vir_h_observation}
\end{align}

It can be seen that the equations are composed of a state equation and a observation equation,
we can introduce KF again to track the channel.
% {Correspondingly,} the  KF process to track  the UL virtual channel   $[\tilde{\mathbf h}_{k,m}]_{\mathcal Q_k}$ can be listed as
%\begin{align}
%&\hat{\mathbf r}_{m_{\mathcal Q_k}}^- = \boldsymbol \alpha^{*}\hat{\mathbf r}_{{m-1}_{\mathcal Q_k}},\\
%&\mathbf \Gamma_{k,m}^{U\!L -} = \boldsymbol \alpha^{*} \mathbf \Gamma_{k,m-1}^{U\!L -} {\boldsymbol \alpha^{*}}^H + \mathbf \Lambda^{*},\\
%&\mathbf K_{k,m}^{U\!L} =  \mathbf \Gamma_{k,m}^{U\!L -} \mathbf D_{\mathcal Q}^H \Big(\mathbf D_{\mathcal Q} \mathbf \Gamma_{k,m}^{U\!L -} \mathbf D_{\mathcal B}^H + \frac{\sigma_n^2 }{\sigma_p^2 }\mathbf I \Big)^{-1},\\
%&\hat{\mathbf r}_{m_{\mathcal Q_k}} = \hat{\mathbf r}_{m_{\mathcal Q_k}}^- \! + \!\mathbf K_{k,m}^{U\!L}\Big(\mathbf y_{g,m} -  \mathbf D_{\mathcal Q} \hat{\mathbf r}_{m_{\mathcal Q_k}}^- \Big),\\
%&\mathbf \Gamma_{k,m}^{U\!L  }  = \Big( \mathbf I - \mathbf K_{k,m}^{U\!L} \mathbf D_{\mathcal Q} \Big)\mathbf \Gamma_{k,m}^{U\!L -}.
%\end{align}

\section{Downlink Channel Model Reconstruction and Channel Restoration}

Similar to \eqref{eq:h_k}, the physical DL channel from the BS to the user $k$ during time block $m$ can be written as:
\begin{align}
  \mathbf g_{k,m} = \int_{-\infty}^{+\infty} \sum_{l=1}^{L} \mathbf a (\varphi_{k,l,m}) e^{\jmath 2\pi \nu mL_cT_s}
    \hbar_{k}(\varphi_{k,l,m},\nu) d\nu,\label{eq:g_k}
\end{align}
where $\varphi$ is the direction of departure (DOD) the propagation path;
$ \mathbf a(\varphi)$ is the BS antenna array spatial steering vector defined in \eqref{steer_vec},
but with different DL carrier wavelength ${\lambda}'$ if FDD mode is selected.
Similar to \eqref{vcr_h}, the DL channel $\mathbf g_{k,m}$ can be also approximated by
 {the} sparse virtual channel model with spatial signatures $\mathcal Q_k^{\prime}$, i.e.,
\begin{align}
\mathbf g_{k,m} = \underbrace { \{ \mathbf A'^H + \mathbf B'^H \text{diag} (\boldsymbol \rho_k^{'})\} }_{\mathbf \Phi(\boldsymbol \rho_{k}^{'})^H} {\text {diag}}(\mathbf c_{k}^{'}) \mathbf r_{k,m}^{'}
 = [ \mathbf \Phi( \boldsymbol \rho_{k}^{'}) ]_{:,\mathcal Q_k^{'}}^H [\mathbf r_{k,m}^{'}]_{\mathcal Q_k^{'}}.
\end{align}
\subsection{DL channel model parameters reconstruction} \label{sec_dl_track}

{In the FDD mode, since the channel covariance {matrices} between UL and DL {have} no reciprocity, the DL model parameters $\boldsymbol\Xi_k^\prime=\{\boldsymbol \rho_k^\prime, \mathbf c_k^\prime, \alpha_k^\prime,\boldsymbol\Lambda_k^\prime, \sigma_{n,k}^{2\prime} \}$ are different from the UL ones.
Thanks to the angle reciprocity, we can reconstruct some parameters in $\boldsymbol\Xi_k^\prime$. However,
$\boldsymbol\Lambda_k^\prime$, $\sigma_{n,k}^{2\prime}$ are closely related with the carrier frequency,
and can not be perfectly inferred from the UL.
In an easy way, an alternative method is to learn those parameters again in the DL training to obtain the model parameters, which will need some dedicated training and will waste the system bandwidth.
Thus,  we will resort to the Bayesian Kalman filtering to implement both the effective channel tracking and the restoration of the model parameters. We will see that this method does not dedicated training period, and will ensure the  real-time channel updating. In the following, we will first introduce the
reconstruction of $\boldsymbol \rho_k^\prime$, $\mathbf c_k^\prime$, $\alpha_k^\prime$.
Then, in the next subsection, the optimal Bayesian Kalman filtering will be given. }

%which will bring an unbearable scale of overhead.
%Luckily, they can be coarsely derived from the UL ones, by which the overhead can be saved and the precision accuracy is in a acceptable range.
%In the following, we will carefully discuss the parameters in $\boldsymbol\Xi_k^\prime$ one by one.

\subsubsection{$\alpha_k^\prime$}
For a specific user, the moving velocities along the UL and DL are the same.
Thus, the Doppler frequency $\nu_k^{\max\prime}$ along the DL can be derived from the known parameters $\lambda$, $\lambda^\prime$ and $\nu_k^{\max}$ as
$\nu_k^{\max\prime} = \frac{\lambda^\prime}{\lambda}\nu_k^{\max}$. Then, $\alpha_k^\prime$ is given by
\begin{align}\label{DL_alpha}
\alpha_k^\prime = J_0(2\pi\nu_k^{\max\prime}L_cT_s).
\end{align}

\subsubsection{$\mathcal Q_k^{\prime}$ {\rm and} $\boldsymbol \rho_k^{\prime} $}
As there's reciprocity lying in the propagation paths of the radiowaves,
it can be found that only the DL signal waves that reverse the UL paths can reach the user in the DL transmission period  \cite{data2001prediction, metis6wireless}.
Hence, the DODs of DL scattering rays is the same as the DOAs of UL radiowaves at the BS.
Therefore, we can recover $\mathcal Q_k^{\prime}$ as well as $\boldsymbol \rho_k^{\prime} $ from $\mathcal Q_k$ and $\boldsymbol \rho_k $.
{Similar to}  \eqref{eq:nonzero_Q}, we have
\begin{align}
\sin(\theta_k) = \frac{(p + [\boldsymbol \rho_k]_p)\lambda}{N_t d} = \frac{(p' + [\boldsymbol \rho_k']_{p'})\lambda'}{N_t d} = \sin(\varphi_k),
p\in \mathcal Q_k.
\end{align}
Then, it can be obtained that
\begin{align} \label{angle reciprocity}
  (p' + [\boldsymbol \rho_k']_{p'}) = \frac{(p + [\boldsymbol \rho_k]_p) \lambda}{\lambda^\prime}, p\in \mathcal Q_k,
\end{align}
where
\begin{align}\label{p' and rho}
  p' = \lfloor \frac{(p + [\boldsymbol \rho_k]_p) \lambda}{\lambda^\prime} \rfloor , p\in \mathcal Q_k,\kern 30pt  [\boldsymbol \rho_k']_{p'} = \frac{(p + [\boldsymbol \rho_k]_p) \lambda}{\lambda^\prime} - p',
\end{align}
{and} $\mathcal Q_k^{\prime} $ includes all the $p' $ that satisfies \eqref{angle reciprocity}.

Notice that different $p \in \mathcal Q $ may be mapped on a same grid in the DL virtual channel.
If two rays in the UL are mapped on a same $p\prime $ with different bias $\rho^\prime $, our scheme is to see them as one ray and adopt the average of their bias.
For example, if the bias of two specific ray is $0.1 $ and $0.3 $, respectively, we regard them as the very ray with the bias $0.2 $.
Furthermore, the corresponding $c_k^\prime$ can be determined by $\mathcal Q_k^\prime$, as $[\mathbf c_k^\prime]_i = 1$ when $i\in \mathcal Q_k^\prime $, .
%\begin{align}\label{DL c}
%[\mathbf c_k^\prime]_i=
%\left\{
%\begin{aligned}
%1~~~& i\in \mathcal Q_k',\\
%0~~~& \text{else.}
%\end{aligned}
%\right.
%\end{align}

\subsection{DL channel restoration by optimal Bayesian Kalman filtering}

{Now, we start to track  $[\tilde{\mathbf g}_{k,m}]_{\mathcal Q_k^{\prime}}$
with the reconstructed partial knowledge about $\boldsymbol\Xi_k^\prime$
in the previous subsection. }
Similar to \eqref{grouping_UL},  the $K$ users is divided	into $G'$ groups  such that the DL spatial signatures of the users in the same group do not overlap, i.e.,
\begin{align}\label{grouping_DL}
\mathbf c_k^{\prime}\mathbf c_j^{\prime T} =0, k\neq j.
\end{align}

Then, the user  {indices} of  {the} group $g$ are collected into  {the} set $\mathcal G_g^\prime$.
In order to avoid the inter-group interference,  the DL channels  for each group are separately estimated.   The training sequences can be reused by  {the}
users in the same group due to the separation of their spatial signatures.
Thus, $|\mathcal Q'_k|$ orthogonal training sequences are required to estimate $|\mathcal Q'_k|$ coefficients for each user.
So we build a $M_g \times M_g $ matrix $\mathbf T_g $ with $\mathbf T_g  \mathbf T_g ^H = M_g \sigma_p \mathbf I_{M_g}(M_g = \max \limits_{k \in \mathcal G'_g} |\mathcal Q'_k|)$
and select $|\mathcal Q'_k|$ rows of $\mathbf T_g $ as the training sequences for user $k$, i.e. $\mathbf S_k = [\mathbf T_g ]_{1:| {{\mathcal Q}_k^\prime }|,:}$.
Then, $\mathbf S_k$ is transmitted on the beam $[\mathbf \Phi(\boldsymbol \rho_k')^H]_{:,\mathcal Q_k^{\prime}}$. Since the BS simultaneously transmits training sequences  for   users in the same group, the transmitted signals during DL channel estimation for group $g$ is given by
$\mathbf \Gamma_g = \sum_{k \in \mathcal G_g^{\prime}}   [\mathbf \Phi(\boldsymbol \rho_k')^H]_{:,\mathcal Q_k^{\prime}} \mathbf S_k $.

% Then, $\mathbf S_k$ is transmitted on the beam $[\mathbf \Phi()^H]_{:,\mathcal Q_i^{\prime}}$. Since the BS simultaneously transmits training sequences  for   users in the same group, the transmitted signals during DL channel estimation for group $g$ is given by
%\begin{align}
%\mathbf \Gamma_g = \sum_{k \in \mathcal G_g^{\prime}}   [\mathbf F^H]_{:,\mathcal Q_k^{\prime}} \mathbf S_k.
%\end{align}

As a result, the received signal at the user $k$ of the group $g$ can be expressed as
\begin{align}
\mathbf y^{\prime H}_{k,m} = \mathbf g^H_{k,m} \mathbf \Gamma_g + \mathbf n_{k,m}^{\prime H}
%&= [\tilde{\mathbf g}_{k,m}]_{\mathcal Q_k^{\prime}}^H  [\mathbf \Phi(\boldsymbol \rho_k')]_{\mathcal Q_k^{\prime},:} [\mathbf \Phi(\boldsymbol \rho_k')]_{:,\mathcal Q_k^{\prime}}^H  \mathbf S_k
%+ \sum_{i \in \mathcal G_l,i\neq k}  [\tilde{\mathbf g}_{k,m}]_{\mathcal Q_k^{\prime}}^H  [\mathbf \Phi(\boldsymbol \rho_k')]_{\mathcal Q_k^{\prime},:} [\mathbf \Phi(\boldsymbol \rho_k')]_{:,\mathcal Q_k^{\prime}}^H  \mathbf S_i
%+ \mathbf n_{k,m}^{\prime H}, \notag \\
= [\tilde{\mathbf g}_{k,m}]_{\mathcal Q_k^{\prime}}^H \mathbf S_k + \mathbf n_{k,m}^{\prime H}.
\end{align}

To eliminate the inter-group interference, we can further derive that
\begin{align}
	\tilde { \mathbf y}_{k,m}^{\prime}   = \frac{1}{ M_g\sigma_p^2} \mathbf S_k \mathbf y_{k,m}^{\prime}
%= \frac{1}{M_g \sigma_p^2}  \mathbf S_k \mathbf S_k^H [\tilde{\mathbf g}_{k,m}]_{\mathcal Q_k^{\prime}}
%+ \frac{1}{ M_g \sigma_p^2} \mathbf S_k \mathbf n_{k,m}^{\prime}  \notag\\
= [\tilde{\mathbf g}_{k,m}]_{\mathcal Q_k^{\prime}}  +   \tilde {\mathbf n}_{k,m}^{\prime},
\end{align}
where  the equivalent Gaussian white noise vector
{$\tilde{\mathbf n}_{k,m}^{\prime} = \frac{1}{ M_g\sigma_p^2} \mathbf S_k \mathbf n_{k,m}^{\prime} \sim \mathcal {CN} (\mathbf 0, \frac{\sigma_n^{2\prime}}{\sigma_p^2} \mathbf I_{|\mathcal Q_{k}^\prime|})$.}
Here the covariance of original noise $\sigma_n^{2\prime} $ is unknown.

Then we can obtain the following state-space model as
\begin{align}
\left\{
\begin{aligned}
\left[\tilde{\mathbf g}_{k,m}^{\prime}\right]_{\mathcal Q_k^{\prime}} &= \alpha_{k}^\prime \left[\tilde{\mathbf g}_{k,m}\right]_{\mathcal Q_k^{\prime}} + \left[{\mathbf v}_{k,m}^{\prime}\right]_{\mathcal Q_k^{\prime}}, \\
\left[\tilde{\mathbf y}_{k,m}^{\prime}\right]_{\mathcal Q_k^{\prime}} &= \left[\tilde{\mathbf g}_{k,m}\right]_{\mathcal Q_k^{\prime}} + \left[\tilde{{\mathbf n}}_{k,m}^{\prime}\right]_{\mathcal Q_k^{\prime}}.
\end{aligned}
\right. \label{eq:AR_downlink}
\end{align}
where $[{\mathbf v}_{k,m}^{\prime}]_{\mathcal Q_k^{\prime}} \sim \mathcal {CN} (0, [\mathbf \Lambda_k^\prime]_{\mathcal Q_k^\prime})$.
{As mentioned in the previous subsection, we can reconstruct partial knowledge about the model parameters in
(\ref{eq:AR_downlink}). However, the statistics of the noise in both the observation and
the state equations are unknown. Thus,
%It's unfortunate that $[\mathbf \Lambda_k^\prime]_{\mathcal Q_k^\prime} $ is dependent on the channel spatial covariance matrix,
%which relies on the carrier frequency, and is very hard to obtain \cite{cov_estimation_FDD}.
%Furthermore, if we obtain the DL channel covariance in a complex way,
%the feedback overhead will also be unaffordable.
%In addition, the method to obtain $[\mathbf \Lambda_k^\prime]_{\mathcal Q_k^\prime} $ in \cite{Ma_J_SBL_Time_varing} is just an empirical inference that can't be confirmed correctly in theory.
%Not only the process noise covariance, the covariance of observation noise also has a unknown parameter $\sigma_n^{2\prime} $.
it is unable to track the DL channel by using the classical KF method, whose performance
is very sensitive to the accuracy of noise statistics.
Nonetheless, there are many robust KF methods to handle this problem, such as IBF KF in\cite{IBR_KF}.
In order to fully utilize the additional information
in the observed signal, the optimal Bayesian Kalman filter (OBKF) method
{will be adopted} for our DL channel tracking process.
The process is divided into 3 parts: the OBKF process, the sum-product algorithm for posterior noise statistics, and the MCMC computation.}

\subsubsection{\textbf{OBKF for DL channel tracking }}

For one specific user, we denote $\boldsymbol \vartheta = \left\{ \sigma_n^{2\prime}, [\mathbf \Lambda^\prime]_{j,j}, j \in \mathcal Q^\prime \right\} $
as the set of all the unknown parameters in
both the process noise and the observation noise vectors,
and use {the} superscript $\boldsymbol\vartheta$
to express that the unknown parameters relate partly or wholly with it.
{Then, the state-space model (\ref{eq:AR_downlink}) can be reexpressed as}
\begin{align}
\left\{
\begin{aligned}
\left[\tilde{\mathbf g}_{m}\right]_{\mathcal Q^{\prime}}^{\boldsymbol \vartheta} = &\alpha^\prime \left[\tilde{\mathbf g}_{m}\right]_{\mathcal Q^{\prime}}^{\boldsymbol \vartheta}
+ \left[{\mathbf v}_{m}^{\prime}\right]_{\mathcal Q^{\prime}}^{\boldsymbol \vartheta},  \\
%%%%%%%%%%%%%%%%%%%%%%%%%%%%%%%%%%%%%%%%%%%%%%%%%%
\left[\tilde{\mathbf y}_{m}^{\prime }\right]_{\mathcal Q^{\prime}}^{\boldsymbol \vartheta} =& \left[\tilde{\mathbf g}_{m}\right]_{\mathcal Q^{\prime}}^{\boldsymbol \vartheta}
+ \left[\tilde{{\mathbf n}}_{m}^{\prime }\right]_{\mathcal Q_k^{\prime}}^{\boldsymbol \vartheta},
\end{aligned}
\right. \label{eq:AR_downlink_final}
\end{align}
{Since each user can track the simultaneously channels and restore
the model parameters independently, we will ignore
the subscript $k$ in the following for simplicity.}

Thus, {under the OBKF framework, the following equations
can be utilized to effectively track the DL virtual channel $[\tilde{\mathbf g}_{k,m}]_{\mathcal Q_k^{\prime}}$ as}
\begin{align}
  &\tilde{\mathbf z}_{m}^{\boldsymbol \vartheta} = \tilde{\mathbf y}_{m}^{\prime {\boldsymbol \vartheta}}
   - [\widehat {\tilde{\mathbf g}}_{m}]_{\mathcal Q^{\prime}}^{\boldsymbol \vartheta} \label{OBKF 1} \\
   %%%%%%%%%%%%%%%%%%%%%%%%%%%%%%%%%%%
   &\mathbf K_{m}^{\Theta} = \mathbb E_{\boldsymbol \vartheta} [\mathbf P_{m}^{\boldsymbol \vartheta} | \tilde{\mathbf y}^{\prime} (m-1)]
   \mathbb E_{\boldsymbol \vartheta}^{-1} \left\{ \mathbf P_{m}^{\boldsymbol \vartheta} + \frac{\sigma_n^{2\prime}}{\sigma_p^2} \mathbf I_{|\mathcal Q_{k}^\prime|} | \tilde{\mathbf y}^{\prime} (m-1) \right\}, \label{OBKF 2} \\
   %%%%%%%%%%%%%%%%%%%%%%%%%%%%%%%%%%%
   &[\widehat {\tilde{\mathbf g}}_{{m+1}}]_{\mathcal Q^{\prime}}^{\boldsymbol \vartheta}
   = \alpha^\prime [\widehat {\tilde{\mathbf g}}_{m}]_{\mathcal Q^{\prime}}^{\boldsymbol \vartheta}
   + \alpha^\prime \mathbf K_{m}^{\Theta} \tilde{\mathbf z}_{m}^{\boldsymbol \vartheta}, \label{OBKF 3} \\
   %%%%%%%%%%%%%%%%%%%%%%%%%%%%%%%%%%%
   &\mathbb E_{\boldsymbol \vartheta} \left\{\mathbf P_{m+1}^{\boldsymbol \vartheta} | \tilde{\mathbf y}^{\prime} (m)\right\}
   = \alpha^{\prime2} (\mathbf I - \mathbf K_{m}^{\Theta} \mathbf )
   \mathbb E_{\boldsymbol \vartheta} \left\{\mathbf P_m^{\boldsymbol \vartheta} | \tilde{\mathbf y}^{\prime} (m)\right\}
   + \mathbb E_{\boldsymbol \vartheta}  \Big\{ [\mathbf \Lambda^{\prime}]_{\mathcal Q^{\prime}} | \tilde{\mathbf y}^{\prime} (m)  \Big\}, \label{OBKF 4}
\end{align}
where $\tilde{\mathbf y}^{\prime} (m) = \left[\tilde{\mathbf y}_{1}^{\prime H}, \tilde{\mathbf y}_{2}^{\prime H}, \ldots, \tilde{\mathbf y}_{m}^{\prime H}  \right]^H $,
and $\mathbf P_{m}^{\boldsymbol \vartheta} = \mathbb E\left\{([{\tilde{\mathbf g}}_{m}]_{\mathcal Q^{\prime}}^{\boldsymbol \vartheta} - [\widehat {\tilde{\mathbf g}}_{m}]_{\mathcal Q^{\prime}}^{\boldsymbol \vartheta})
([ {\tilde{\mathbf g}}_{m}]_{\mathcal Q^{\prime}}^{\boldsymbol \vartheta} - [\widehat {\tilde{\mathbf g}}_{m}]_{\mathcal Q^{\prime}}^{\boldsymbol \vartheta})^H\right\} $ is {the}
covariance matrix {of the channel estimation error}  relative to ${\boldsymbol \vartheta}$
at time $m$.

%To obtain $\mathbb E_{\boldsymbol \vartheta} \left\{\mathbf P_{m}^{\boldsymbol \vartheta} | \tilde{\mathbf y}^{\prime} (m)\right\} $
%there are two options.
%The first option is to do the recursive updates in \eqref{OBKF 2}-\eqref{OBKF 4} at each time block $m $
%from the beginning using $\mathbb E_\theta [\mathbf Q_k^{\theta_1} | \mathcal Y_{k,m}'] $ and $\mathbb E_\theta [\mathbf R_k^{\theta_2} | \mathcal Y_{k,m}'] $,
%i.e., first compute $\mathbf K_{k,0}^{\Theta *} $
%using $\text{cov} [[{\tilde{\mathbf g}}_{k,{m+1}}]_{\mathcal Q_k^{\prime}}^{\theta}]$ and $\mathbb E_\theta [\mathbf R_k^{\theta_2} | \mathcal Y_{k,m}'] $,
%then compute $\mathbb E_\theta [\mathbf P_{k,1}^{{\tilde{\mathbf g}}, \theta} | \mathcal Y_{k,m}'] $
%using $\mathbf K_{k,0}^{\Theta *} $ and $\mathbb E_\theta [\mathbf Q_k^{\theta_1} | \mathcal Y_{k,m}'] $,
%then compute $\mathbf K_{k,1}^{\Theta *} $
%using $\mathbb E_\theta [\mathbf P_{k,1}^{{\tilde{\mathbf g}}, \theta} | \mathcal Y_{k,m}'] $ and $\mathbb E_\theta [\mathbf R_k^{\theta_2} | \mathcal Y_{k,m}'] $,
%and so on, until we reach $\mathbb E_\theta [\mathbf P_{k,m}^{{\tilde{\mathbf g}}, \theta} | \mathcal Y_{k,m}'] $.
%The other option is to
{To decrease the computation complexity,}
we make the approximation
$\mathbb E_{\boldsymbol \vartheta} \!\!\left\{\!\mathbf P_{m}^{\boldsymbol \vartheta} \!| \tilde{\mathbf y}^{\prime} \!(\!m\!)\!\right\} \!\!\approx\!\! \mathbb E_{\boldsymbol \vartheta} \!\!\left\{\!\mathbf P_{m}^{\boldsymbol \vartheta} | \tilde{\mathbf y}^{\prime} \!(\!m\!-\!1\!)\! \right\} $ for simplicity,
and replace {$\mathbb E_{\boldsymbol \vartheta} \left\{\mathbf P_{m}^{\boldsymbol \vartheta} | \tilde{\mathbf y}^{\prime} (m)\right\} $} in \eqref{OBKF 4} with $\mathbb E_{\boldsymbol \vartheta} \left\{\mathbf P_{m}^{\boldsymbol \vartheta} | \tilde{\mathbf y}^{\prime} (m-1)\right\} $ {from the previous iteration} \cite{OBKF}.
This option is computationally more efficient,
{which is due to the fact that}
we do not need to repeat all the recursions in (\ref{OBKF 1})--(\ref{OBKF 4}) at each time block $m $.
%We used the second option in our simulations.

From \eqref{OBKF 1}, \eqref{OBKF 2}, \eqref{OBKF 3}, \eqref{OBKF 4}, we will find that {two conditional expectations}
$\mathbb E_{\boldsymbol \vartheta} \left\{\frac{\sigma_n^{2\prime}}{\sigma_p^2} \mathbf I_{|\mathcal Q^\prime|} | \tilde{\mathbf y}^{\prime} (m) \right\}$
and $\mathbb E_{\boldsymbol \vartheta} \left\{ [\mathbf \Lambda^{\prime}]_{\mathcal Q^{\prime}} | \tilde{\mathbf y}^{\prime} (m)\right\}$
{should be evaluated with respect to}
the posterior distribution
$p(\!{\boldsymbol \vartheta}\! | \tilde{\mathbf y}^{\prime}\! (\!m\!)) \!\propto\! p(\tilde{\mathbf y}^{\prime} \!(\!m\!)\! | {\boldsymbol \vartheta}) p(\!{\boldsymbol \vartheta}\!) $,
where $p(\tilde{\mathbf y}^{\prime} (m) | {\boldsymbol \vartheta}) $ is the likelihood function of ${\boldsymbol \vartheta} $ given the observation sequence $\tilde{\mathbf y}^{\prime} (m) $.
{Since} there {may be} no closed-form solution for $ p ({\boldsymbol \vartheta} | \tilde{\mathbf y}^{\prime} (m)) $ for many prior distributions,
to implement the OBKF process,
we employ {the} MCMC method to generate samples from the posterior distribution $ p ({\boldsymbol \vartheta} | \tilde{\mathbf y}^{\prime} (m)) $
and {to} approximate $\mathbb E_{\boldsymbol \vartheta} \left\{\frac{\sigma_n^{2\prime}}{\sigma_p^2} \mathbf I_{|\mathcal Q^\prime|} | \tilde{\mathbf y}^{\prime} (m) \right\}$
and $\mathbb E_{\boldsymbol \vartheta} \left\{[\mathbf \Lambda^{\prime}]_{\mathcal Q^{\prime}} | \tilde{\mathbf y}^{\prime} (m)\right\} $
as sample means of the generated MCMC samples.
With the Bayes rule, it can be checked
that {the likelihood function $p (\tilde{\mathbf y}^{\prime} (m) | {\boldsymbol \vartheta}) $
should be calculated to determine  $ p ({\boldsymbol \vartheta} | \tilde{\mathbf y}^{\prime} (m)) $}.

{With  (\ref{eq:AR_downlink_final})
and the property of the the Markov  model},
we can obtain
 %in the state-space model postulates that,
%given the current state $[\tilde{\mathbf g}_{m}]_{\mathcal Q^{\prime}}^{\boldsymbol \vartheta} $,
%the next state $[\tilde{\mathbf g}_{m+1}]_{\mathcal Q^{\prime}}^{\boldsymbol \vartheta} $
%and the current observation $\tilde{\mathbf y}_{m}^\prime $
%are normally distributed and possess the following conditional independence properties:
\begin{align}
  p(\tilde{\mathbf y}_{m}^\prime | \tilde{\mathbf y}^{\prime} (m-1), \mathbf x^\prime (m); {\boldsymbol \vartheta})
  &=  p(\tilde{\mathbf y}_{m}^\prime | [\tilde{\mathbf g}_{m}]_{\mathcal Q^{\prime}}; \boldsymbol \vartheta)
  = \mathcal {CN} \left(\tilde{\mathbf y}_{m}^\prime;  \mathbf [\tilde{\mathbf g}_{m}]_{\mathcal Q^{\prime}}, \frac{\sigma_n^{2\prime}}{\sigma_p^2} \mathbf I_{|\mathcal Q^\prime|} \right), \label{y' PDF} \\
  %%%%%%%%%%%%%%%%%%%%%%%%%%%%%%%%%%%%%%%%%%%%%%%%%%%%%%%%%%%%%%%%%%%%%%%%%%%%
  p(\mathbf [\tilde{\mathbf g}_{m+1}]_{\mathcal Q^{\prime}} | \tilde{\mathbf y}^{\prime} (m+1), \mathbf x^\prime (m); {\boldsymbol \vartheta})
  &= p(\mathbf [\tilde{\mathbf g}_{m+1}]_{\mathcal Q^{\prime}} | \mathbf [\tilde{\mathbf g}_{m}]_{\mathcal Q^{\prime}}; {\boldsymbol \vartheta})
  = \mathcal {CN} \left([\tilde{\mathbf g}_{m+1}]_{\mathcal Q^{\prime}}; \alpha^\prime [\tilde{\mathbf g}_{m}]_{\mathcal Q^{\prime}}, [\mathbf \Lambda^{\prime}]_{\mathcal Q^{\prime}} \right), \label{g_m+1 PDF}
\end{align}
where $\mathbf x^\prime (m) = [[\tilde{\mathbf g}_{1}]_{\mathcal Q^{\prime}}^H, [\tilde{\mathbf g}_{2}]_{\mathcal Q^{\prime}}^H, \ldots, [\tilde{\mathbf g}_{m}]_{\mathcal Q^{\prime}}^H ]^H $ is the set of the past $m$ $[\tilde{\mathbf g}_{m}]_{\mathcal Q^{\prime}} $.

{With (\ref{y' PDF}) and (\ref{g_m+1 PDF}),the marginalization of $p(\tilde{\mathbf y}^{\prime} (m), \mathbf x^\prime (m) | {\boldsymbol \vartheta}) $
can be factorized as}
\begin{align}\label{y marginalization PDF}
  &p(\tilde{\mathbf y}^{\prime} (m)  | {\boldsymbol \vartheta}) = \underbrace{\int \dots \int }_{[\tilde{\mathbf g}_{1}]_{\mathcal Q^{\prime}}, \ldots, [\tilde{\mathbf g}_{m}]_{\mathcal Q^{\prime}}}
  p(\tilde{\mathbf y}^{\prime} (m-1) , \mathbf x^\prime (m) | {\boldsymbol \vartheta})d[\tilde{\mathbf g}_{1}]_{\mathcal Q^{\prime}}, \ldots, d[\tilde{\mathbf g}_{m}]_{\mathcal Q^{\prime}} \notag \\
  &= \underbrace{\int \dots \int }_{[\tilde{\mathbf g}_{1}]_{\mathcal Q^{\prime}}, \ldots, [\tilde{\mathbf g}_{m}]_{\mathcal Q^{\prime}}}
  \prod_{i=1}^{m} p(\tilde{\mathbf y}_{i}^\prime | [\tilde{\mathbf g}_{i}]_{\mathcal Q^{\prime}}, {\boldsymbol \vartheta})
  \prod_{i=1}^{m} p(\mathbf [\tilde{\mathbf g}_{i}]_{\mathcal Q^{\prime}} | \mathbf [\tilde{\mathbf g}_{i-1}]_{\mathcal Q^{\prime}}, {\boldsymbol \vartheta})
  p([\tilde{\mathbf g}_{1}]_{\mathcal Q^{\prime}})
  d[\tilde{\mathbf g}_{1}]_{\mathcal Q^{\prime}}, \ldots, d[\tilde{\mathbf g}_{m}]_{\mathcal Q^{\prime}}.
\end{align}

Then, $p(\tilde{\mathbf y}^{\prime} (m)  | {\boldsymbol \vartheta})$ can be denoted with a factor
graph, as shown in \figurename{ \ref{factor graph}},
where the factors in (\ref{y marginalization PDF}) are represented by
\textquotedblleft functions nodes \textquotedblright marked {blue and red} boxes
and the corresponding random variables are represented by
\textquotedblleft variable nodes\textquotedblright marked as {green} circles.
%Obviously, there are four kinds of function nodes, i.e.,
%$f^A_{m,p}$,
%$f^B_{m,i}$,
%$f^C_{m,i}$, and
%$f^D_i$, and three types of variable nodes, i.e., $\tilde h_{k,m,i}$, $r_{k,m,i}$, $c_{k,i}$, in \textcolor{blue}{\{Please add one figure here\}}.
 One specific variable node $\boldsymbol x$ connects with
 the function nodes $f$, whose augments contain $\boldsymbol x$.
 Furthermore, we will resort to the belief propagation (BP), also known as sum-product message passing,
 to implement the message-passing in our constructed factor graph \figurename{ \ref{factor graph}}. BP passes real valued messages along
 the edges between nodes in the factor graph. Specifically, for
 the function node $f$ and the variable node $x$, the messages from $f$ to $\boldsymbol x$
 and from $\boldsymbol x$ to $f$ are separately defined as $\Omega_{f\rightarrow \boldsymbol x}(x)$ and
 $\Omega_{\boldsymbol x\rightarrow f}(\boldsymbol x)$, whose augment is $\boldsymbol x$. With
 the BP theory, we can obtain
 \begin{align}
 \Omega_{\boldsymbol x\rightarrow f}(\boldsymbol x)=\prod_{f^\prime\in \mathcal N(\boldsymbol x)\slash f} \Omega_{f^\prime \rightarrow \boldsymbol x}(\boldsymbol x),\kern 20pt
 \Omega_{f\rightarrow \boldsymbol x}(\boldsymbol x)=\int_{\sim \boldsymbol x}\Big(f(\boldsymbol x)\prod_{\boldsymbol x^\prime\in\mathcal{N}(f)\slash \boldsymbol x}\Omega_{\boldsymbol x^\prime\rightarrow f}(x^\prime))\Big),\label{eq:sum_pro_rule}
 \end{align}
where the set $\mathcal N(\boldsymbol x)$ collects all the neighbouring nodes of the given node $\boldsymbol x$
in one factor graph, and $\sim \boldsymbol x$ possesses the same meaning with the same notation \cite{factorgraph}.
%
%\eqref{y marginalization PDF} can be regarded as the factorization of a global function for which we can employ the sum-product algorithm,
%which is a message-passing algorithm.
%To visualize the computations in a sum-product algorithm, we introduce a factor graph about it.

\subsubsection{\textbf{Sum-Product Algorithm for posterior noise statistics}}

\begin{figure}
  \centering
  \includegraphics[width=160mm]{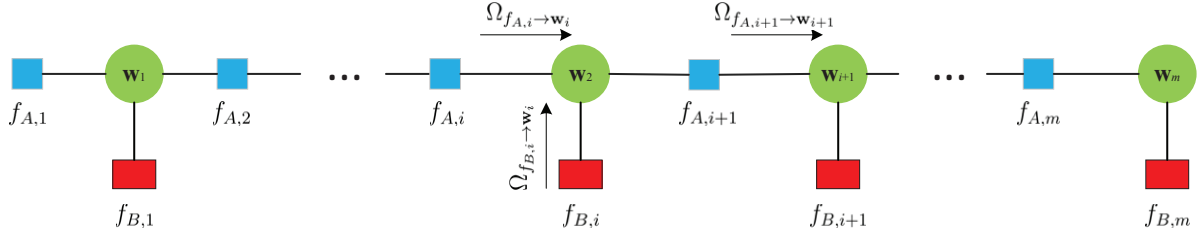}\\
  \caption{factor graph model to calculate $f(\tilde{\mathbf y}^{\prime} (m) | \boldsymbol \vartheta)$. }\label{factor graph}
\end{figure}

A node in the factor graph operates when it receives all messages from its {neighbouring} nodes.
The first step {to} run a factor graph is that each leaf function node sends {the message} to its {neighbouring} nodes.
For expression simplicity, we define the factor nodes and variable nodes in \figurename{ \ref{factor graph}} as
\begin{align}\label{nodes}
  \mathbf w_i &= [\tilde{\mathbf g}_{i}]_{\mathcal Q^{\prime}}, \notag \\
  f_{A,i} &= \mathcal {CN} \left(\mathbf w_i; \alpha^\prime \mathbf w_{i-1}, [\mathbf \Lambda^{\prime}]_{\mathcal Q^{\prime}} \right),
  f_{B,i} = \mathcal {CN} \left(\tilde{\mathbf y}_{i}^\prime;  \mathbf w_i, \frac{\sigma_n^{2\prime}}{\sigma_p^2} \mathbf I_{|\mathcal Q^\prime|} \right), i = 1, \dots, m,
\end{align}
{and} $f_{A,1} = p(\mathbf w_1) = \mathcal{CN} \left(\mathbf w_1; \mathbf 0, [\mathbf \Lambda^{\prime}]_{\mathcal Q^{\prime}} \right) $.

%Then we draw the factor graph corresponding to our objective function \eqref{y marginalization PDF}.

%\figurename{ \ref{factor graph}} shows the factor graph model adopted to compute the function in \eqref{y marginalization PDF}.
It can be seen {from \figurename{ \ref{factor graph}}}
that there's three kind of message in the factor graph, i.e., $\Omega_{f_{A,i} \to \mathbf w_i} $, $\Omega_{f_{B,i} \to \mathbf w_i} $, and $\Omega_{\mathbf w_i \to f_{A,i}} $.
{Since}, we only need to consider the forward passing message,
%and the expression of the message $\Omega_{f_{A,i+1} \to \mathbf w_{i+1}}$ is not difficult to express,
the expression of $\Omega_{\mathbf w_i \to f_{A,i}} $ can be omitted here.
With (\ref{y' PDF}), (\ref{g_m+1 PDF}),  and (\ref{eq:sum_pro_rule}),
it can be readily checked from  \figurename{ \ref{factor graph}}  that
\begin{align}
  \Omega_{f_{B,i} \to \mathbf w_i} &= \mathcal {CN} \left(\tilde{\mathbf y}_{i}^\prime;  \mathbf w_i, \frac{\sigma_n^{2\prime}}{\sigma_p^2} \mathbf I_{|\mathcal Q^\prime|} \right) \label{fB to omega}.
  %%%%%%%%%%%%%%%%%%%%%%%%%%%%%%%%%%%%%%%%%%%%%%%%%%%%%%%%
\end{align}
With respect to the term   $\Omega_{f_{A,i} \to \mathbf w_i}$,  we have the following lemma.

\begin{lemma}\label{lemma1}
For all $1\leq i \leq m-1$, the message $\Omega_{f_{A,i+1} \to \mathbf w_{i+1}}$ in \figurename{ \ref{factor graph}} can be expressed as:
\begin{align}\label{Omega A_i+1}
  \Omega_{f_{A,i+1} \to \mathbf w_{i+1}} = \omega_{i+1} \mathcal {CN} \left(\mathbf w_{i+1}; \boldsymbol \mu_{i+1}, \mathbf \Sigma_{i+1} \right),
\end{align}
where
\begin{align}
  \boldsymbol\Sigma_{i+1}=&\left([\bm\Lambda^\prime]_{\mathcal Q^\prime}^{-1}-\alpha^{\prime2}[\bm\Lambda^\prime]_{\mathcal Q^\prime}^{-1}\bm\Gamma_i
   [\bm\Lambda^\prime]_{\mathcal Q^\prime}^{-1}\right)^{-1},\\
   %%%%%%%%%%%%%%%%%%%%%%%%%%%%%%%%%%%%%%%%%%%%%%%%%%%%%%%%%%%%
   \bm\mu_{i+1}=& \alpha^\prime \bm\Sigma_{i+1}[\bm\Lambda^\prime]_{\mathcal Q^\prime}^{-1}\bm\Gamma_i
   \left(\left(\frac{\sigma_n^{2\prime}}{\sigma_p^2} \mathbf I_{|\mathcal Q^\prime|}\right)^{-1}
   \tilde{\mathbf y}_{i}^\prime + \mathbf \Sigma_{i}^{-1} \boldsymbol \mu_{i}\right),
  \end{align}
  \begin{small}
  \begin{align}\label{omega_i+1}
 \omega_{i+1} = \omega_{i} \mathcal{CN}\left(\mathbf 0;\tilde{\mathbf y}_{i}^\prime,
  \frac{\sigma_n^{2\prime}}{\sigma_p^2} \mathbf I_{|\mathcal Q^\prime|}\right)
  \mathcal{CN}\left(\mathbf 0;\bm{\mu}_i,\bm{\Sigma}_i\right)
  \frac{ |\bm\Gamma_i||\bm\Sigma_{i+1}| \cdot
  {|(\bm\Gamma_i+\alpha^{\prime2}\bm\Gamma_i[\bm\Lambda^\prime]_{\mathcal Q^\prime}^{-1}
  \bm\Sigma_{i+1}[\bm\Lambda^\prime]_{\mathcal Q^\prime}^{-1}\bm\Gamma_i)|}}{|[\bm\Lambda^\prime]_{\mathcal Q^\prime}|
  \mathcal{CN}\left(\bm\nu_i,\mathbf 0,(\bm\Gamma_i+\alpha^{\prime2}\bm\Gamma_i[\bm\Lambda^\prime]_{\mathcal Q^\prime}^{-1}
  \bm\Sigma_{i+1}[\bm\Lambda^\prime]_{\mathcal Q^\prime}^{-1}\bm\Gamma_i)^{-1}\right)}.
\end{align}
\end{small}
where $\mathbf \Gamma_i$ and $\bm \nu_{i}$ are defined in the following proof part.
Furthermore, in every step $i$, the parameters
$\omega_{i+1}$, $\boldsymbol \mu_{i+1}$, and $\mathbf \Sigma_{i+1} $ in $ \Omega_{f_{A,i+1} \to \mathbf w_{i+1}}$
are related and only related to those in $ \Omega_{f_{A,i} \to \mathbf w_{i}}$.
In addition, it is {checked} that $\Omega_{f_{A,1} \to \mathbf w_{1} } = \omega_1 \mathcal{CN} \left( \mathbf w_1; \mathbf \mu_1, \mathbf \Sigma_1 \right) $.

\emph{\textbf{Proof}}:\\
Before proceeding, we give the following property:
\begin{align}
\mathcal{CN}\left(\mathbf w_{i+1};\alpha^\prime \mathbf w_i,[\mathbf \Lambda^{\prime}]_{\mathcal Q^{\prime}}\right)
=&\frac{1}{|\pi [\mathbf \Lambda^{\prime}]_{\mathcal Q^{\prime}}|}
\exp\left(-(\mathbf w_{i+1}-\alpha^\prime \mathbf w_i)^H[\mathbf \Lambda^{\prime}]_{\mathcal Q^{\prime}}^{-1}
(\mathbf w_{i+1}-\alpha^\prime \mathbf w_i)\right)\notag\\
%=&\frac{\alpha^{\prime-2|\mathcal Q^\prime|}}{|\pi \alpha^{\prime-2}[\mathbf \Lambda^{\prime}]_{\mathcal Q^{\prime}}|}\exp\left(-(\mathbf w_{i}-\mathbf w_{i+1}/\alpha^\prime)^H[\alpha^{-2}\mathbf \Lambda^{\prime}]_{\mathcal Q^{\prime}}^{-1}
%(\mathbf w_{i}-\mathbf w_{i+1}/\alpha^\prime)\right)\notag\\
=&\alpha^{\prime-2|\mathcal Q^\prime|}\mathcal{CN}\left(\mathbf w_i;\mathbf w_{i+1}/\alpha^\prime,[\alpha^{\prime-2}\mathbf \Lambda^{\prime}]_{\mathcal Q^{\prime}}\right).\label{eq:CSCN_trans_0}
\end{align}

{With the above equation,} if  $\Omega_{f_{A,i} \to \mathbf w_i}=\omega_{i}\mathcal{CN}\left(\mathbf w_i;\boldsymbol\mu_i,\boldsymbol\Sigma_{i}\right)$ holds for $i\ge 1$, we can derive
\begin{align}
  &\Omega_{f_{A,i+1} \to \mathbf w_{i+1}} = \int_{\mathbf w_i} f_{A,i+1} \Omega_{f_{A,i} \to \mathbf w_i} \Omega_{f_{B,i} \to \mathbf w_i} d \mathbf w_i  \notag \\
  %%%%%%%%%%%%%%%%%%%%%%%%%%%%%%%%%%%%%%%%%%%%%%%%%%%%
  &\kern 42pt = \omega_i\int_{\mathbf w_i}
  \mathcal {CN}(\mathbf w_{i+1}; \alpha^\prime \mathbf w_i, [\mathbf \Lambda^{\prime}]_{\mathcal Q^{\prime}})
  \mathcal {CN}\left(\tilde{\mathbf y}_{i}^\prime;  \mathbf w_i, \frac{\sigma_n^{2\prime}}{\sigma_p^2} \mathbf I_{|\mathcal Q^\prime|}\right)\mathcal {CN}(\mathbf w_i; \boldsymbol \mu_{i}, \mathbf \Sigma_{i}) d \mathbf w_i\notag \\
  %%%%%%%%%%%%%%%%%%%%%%%%%%%%%%%%%%%%%%%%%%%%%%%%%%%
%  =& \omega_{i}\alpha^{\prime-2|\mathcal Q^\prime|}\int_{\mathbf w_i}
%  \mathcal{CN}\left(\mathbf w_i;\mathbf w_{i+1}/\alpha^\prime,[\alpha^{-2}\mathbf \Lambda^{\prime}]_{\mathcal Q^{\prime}}\right)
%  \mathcal {CN}\left(\mathbf w_i;  \tilde{\mathbf y}_{i}^\prime, \frac{\sigma_n^{2\prime}}{\sigma_p^2} \mathbf I_{|\mathcal Q^\prime|}\right)
%  \mathcal {CN}(\mathbf w_i; \boldsymbol \mu_{i}, \mathbf \Sigma_{i}) d \mathbf w_i \notag\\
  %%%%%%%%%%%%%%%%%%%%%%%%%%%%%%%%%%%%%%%%%%%%%%%%%%%%
  &\kern 42pt = \omega_{i}\alpha^{\prime-2|\mathcal Q^\prime|}\mathcal{CN}\left(0;\tilde{\mathbf y}_{i}^\prime,
  \frac{\sigma_n^{2\prime}}{\sigma_p^2} \mathbf I_{|\mathcal Q^\prime|}\right)
  \mathcal{CN}\left(\mathbf 0;\bm{\mu}_i,\bm{\Sigma}_i\right)
  \frac{\mathcal{CN}\left(0;\alpha^{\prime-1}\mathbf{w}_{i+1},
  [\alpha^{\prime-2}\mathbf \Lambda^{\prime}]_{\mathcal Q^{\prime}}\right)}
  {\mathcal{CN}\left(\mathbf 0;\bm\Gamma_i
  \left(\alpha^\prime [\mathbf \Lambda^{\prime}]_{\mathcal Q^{\prime}}^{-1}\mathbf{w}_{i+1}
  +\bm\nu_i\right), \bm\Gamma_i\right)},\label{eq:Omega_fa_w_0}
   \end{align}
  where
\begin{align}\label{W Lambda i}
  \bm\nu_i =& \left(\frac{\sigma_n^{2\prime}}{\sigma_p^2} \mathbf I_{|\mathcal Q^\prime|}\right)^{-1} \tilde{\mathbf y}_{i}^\prime + \mathbf \Sigma_{i}^{-1} \boldsymbol \mu_{i},\kern 10pt
  %%%%%%%%%%%%%%%%%%%%%%%%%%%%%%%%%%%%%%%%%%%%%%%%%%%%%%%%
  \mathbf \Gamma_{i}= \left(\alpha^{\prime2} [\mathbf \Lambda^{\prime}]_{\mathcal Q^{\prime}}^{-1} + (\frac{\sigma_n^{2\prime}}{\sigma_p^2} \mathbf I_{|\mathcal Q^\prime|})^{-1} + \mathbf \Sigma_{i}^{-1} \right)^{-1},
\end{align}
{and (\ref{eq:cscn_prod_0}) in the Appendix are utilized
  in the above derivations.}

{Furthermore, with respect to the term} $\frac{\mathcal{CN}\left(0;\alpha^{\prime-1}\mathbf{w}_{i+1},
  [\alpha^{\prime-2}\mathbf \Lambda^{\prime}]_{\mathcal Q^{\prime}}\right)}
  {\mathcal{CN}\left(\mathbf 0;\bm\Gamma_i
  \left(\alpha^\prime[\mathbf \Lambda^{\prime}]_{\mathcal Q^{\prime}}^{-1}\mathbf{w}_{i+1}
  +\bm\nu_i\right), \bm\Gamma_i\right)} $
  in (\ref{eq:Omega_fa_w_0}), we can obtain
    \begin{align}\label{simplify}
  &\frac{\mathcal{CN}\left(0;\alpha^{\prime -1}\mathbf{w}_{i+1},
  [\alpha^{\prime -2}\mathbf \Lambda^{\prime}]_{\mathcal Q^{\prime}}\right)}
  {\mathcal{CN}\left(\mathbf 0;\bm\Gamma_i
  \left(\alpha^\prime [\mathbf \Lambda^{\prime}]_{\mathcal Q^{\prime}}^{-1}\mathbf{w}_{i+1}
  +\bm\nu_i\right),
  \bm\Gamma_i\right)}=
  \frac{\alpha^{\prime 4|\mathcal Q^\prime|}| \bm{\Gamma}_i|^2}{{|[\bm\Lambda^\prime]_{\mathcal Q^\prime}|^2}}
  \frac{\mathcal{CN}(\mathbf w_{i+1};\mathbf 0,[\bm\Lambda^\prime]_{\mathcal Q^\prime})}
  {\mathcal{CN}\left(\mathbf w_{i+1};-\alpha^{\prime-1} [\mathbf \Lambda^{\prime}]_{\mathcal Q^{\prime}}\bm\nu_i,\alpha^{\prime-2} [\mathbf \Lambda^{\prime}]_{\mathcal Q^{\prime}}\bm\Gamma_i^{-1} [\mathbf \Lambda^{\prime}]_{\mathcal Q^{\prime}}\right)}\notag\\
   %%%%%%%%%%%%%%%%%%%%%%%%%%%%%%%%%%%%%%%%%%%%%%%%%%%%%%%%%%%%%%%%%%
   &=\frac{\alpha^{\prime4|\mathcal Q^\prime|}| \bm{\Gamma}_i|^2}{|[\bm\Lambda^\prime]_{\mathcal Q^\prime}|^2}\mathcal{CN}(\mathbf w_{i+1};\bm\mu_{i+1},\bm\Sigma_{i+1})
   \frac{\mathcal{CN}\left(\mathbf 0;\mathbf 0,[\bm\Lambda^\prime]_{\mathcal Q^\prime}\right)}{\mathcal{CN}\left(\mathbf 0;-\alpha^{\prime-1} [\mathbf \Lambda^{\prime}]_{\mathcal Q^{\prime}}\bm\nu_i,\alpha^{\prime-2} [\mathbf \Lambda^{\prime}]_{\mathcal Q^{\prime}}\bm\Gamma_i^{-1} [\mathbf \Lambda^{\prime}]_{\mathcal Q^{\prime}}\right)
   \mathcal{CN}(\mathbf 0;\bm\mu_{i+1},\bm\Sigma_{i+1})}\notag\\
   %%%%%%%%%%%%%%%%%%%%%%%%%%%%%%%%%%%%%%%%%%%%%%%%%%%%%%%%%%%%%%%%%%%
%   &=\frac{\alpha^{\prime4|\mathcal Q^\prime|}| \bm{\Gamma}_i|^2}
%   {|[\bm\Lambda^\prime]_{\mathcal Q^\prime}|^2}
%   \mathcal{CN}(\mathbf w_{i+1};\bm\mu_{i+1},\bm\Sigma_{i+1})
%   \frac{\mathcal{CN}\left(\mathbf 0;\mathbf 0,[\bm\Lambda^\prime]_{\mathcal Q^\prime}\right)}{\mathcal{CN}\left(\bm\nu_i;\mathbf 0,\bm\Gamma^{-1}\right)
%   \mathcal{CN}(\bm\nu_i;\mathbf 0,\alpha^{\prime-2}\bm\Gamma_i^{-1}
%[\bm\Lambda^\prime]_{\mathcal Q^\prime}\bm\Sigma_{i+1}^{-1}
%[\bm\Lambda^\prime]_{\mathcal Q^\prime}\bm\Gamma_i^{-1})}\notag\\
%%%%%%%%%%%%%%%%%%%%%%%%%%%%%%%%%%%%%%%%%%%%%%%%%%%%%%%%%%%%%%%%%%%%%%%%%
%&\kern 10pt \times\frac{|\alpha^{\prime-2} [\mathbf \Lambda^{\prime}]_{\mathcal Q^{\prime}}\bm\Gamma_i^{-1} [\mathbf \Lambda^{\prime}]_{\mathcal Q^{\prime}}|}{|\bm\Gamma_{i}^{-1}|}
%\frac{|\bm\Sigma_{i+1}|}
%{|\alpha^{\prime-2}\bm\Gamma_i^{-1}
%[\bm\Lambda^\prime]_{\mathcal Q^\prime}\bm\Sigma_{i+1}^{-1}
%[\bm\Lambda^\prime]_{\mathcal Q^\prime}\bm\Gamma_i^{-1}|}\notag\\
%%%%%%%%%%%%%%%%%%%%%%%%%%%%%%%%%%%%%%%%%%%%%%%%%%%%%%%%%%%%%%%%%%%%%%%%%
&=\frac{\alpha^{\prime4|\mathcal Q^\prime|}|\bm\Gamma_i|^4|\bm\Sigma_{i+1}|^2}{
|[\bm\Lambda^\prime]_{\mathcal Q^\prime}|^2}
\mathcal{CN}(\mathbf 0;\mathbf 0,[\bm\Lambda^\prime]_{\mathcal Q^\prime})
\frac
{\mathcal{CN}\left(\mathbf 0,\mathbf 0,(\bm\Gamma_i+\alpha^{\prime2}\bm\Gamma_i[\bm\Lambda^\prime]_{\mathcal Q^\prime}^{-1}
  \bm\Sigma_{i+1}[\bm\Lambda^\prime]_{\mathcal Q^\prime}^{-1}\bm\Gamma_i)^{-1}\right)}
{\mathcal{CN}\left(\bm 0;\mathbf 0,\bm\Gamma^{-1}\right)
   \mathcal{CN}(\bm 0;\mathbf 0,
   \alpha^{\prime-2}\bm\Gamma_i^{-1}
[\bm\Lambda^\prime]_{\mathcal Q^
\prime}\bm\Sigma_{i+1}^{-1}
[\bm\Lambda^\prime]_{\mathcal Q^\prime}\bm\Gamma_i^{-1})}\notag\\
%%%%%%%%%%%%%%%%%%%%%%%%%%%%%%%%%%%%%%%%%%%%%%%%%%%%%%%%%%%%%%%%%%%%%%%%%%%%
&\kern10pt\times\frac{1}{\mathcal{CN}\left(\bm\nu_i,\mathbf 0,(\bm\Gamma_i+\alpha^{\prime2}\bm\Gamma_i[\bm\Lambda^\prime]_{\mathcal Q^\prime}^{-1}
  \bm\Sigma_{i+1}[\bm\Lambda^\prime]_{\mathcal Q^\prime}^{-1}\bm\Gamma_i)^{-1}\right)}
  \mathcal{CN}\left(\mathbf w_{i+1},\bm\mu_{i+1},\bm\Sigma_{i+1}\right)\notag\\
  %%%%%%%%%%%%%%%%%%%%%%%%%%%%%%%%%%%%%%%%%%%%%%%%%%%%%%%%%%%%%%%%%%%%%%%%%%
  &=\frac{\alpha^{\prime4|\mathcal Q^\prime|}|\bm\Gamma_i|^4|\bm\Sigma_{i+1}|^2}{
|[\bm\Lambda^\prime]_{\mathcal Q^\prime}|^2}
\frac{1}{|[\bm\Lambda^\prime]_{\mathcal Q^\prime}|}
\frac{|(\bm\Gamma_i+\alpha^{\prime2}\bm\Gamma_i[\bm\Lambda^\prime]_{\mathcal Q^\prime}^{-1}
  \bm\Sigma_{i+1}[\bm\Lambda^\prime]_{\mathcal Q^\prime}^{-1}\bm\Gamma_i)|}{|\bm\Gamma_i
  |\cdot| \alpha^{\prime2}\bm\Gamma_i
[\bm\Lambda^\prime]_{\mathcal Q^
\prime}^{-1}\bm\Sigma_{i+1}
[\bm\Lambda^\prime]_{\mathcal Q^\prime}^{-1}\bm\Gamma_i)|}\notag\\
%%%%%%%%%%%%%%%%%%%%%%%%%%%%%%%%%%%%%%%%%%%%%%%%%%%%%%%%%%%%%%%%%%%%%%%%%%%%
&\kern10pt\times\frac{1}{\mathcal{CN}\left(\bm\nu_i,\mathbf 0,(\bm\Gamma_i+\alpha^{\prime2}\bm\Gamma_i[\bm\Lambda^\prime]_{\mathcal Q^\prime}^{-1}
  \bm\Sigma_{i+1}[\bm\Lambda^\prime]_{\mathcal Q^\prime}^{-1}\bm\Gamma_i)^{-1}\right)}
  \mathcal{CN}\left(\mathbf w_{i+1},\bm\mu_{i+1},\bm\Sigma_{i+1}\right)\notag\\
  %%%%%%%%%%%%%%%%%%%%%%%%%%%%%%%%%%%%%%%%%%%%%%%%%%%%%%%%%%%%%%%%%%%%%%%%%%%
  &=\frac{\alpha^{\prime2|\mathcal Q^\prime|}|\bm\Gamma_i||\bm\Sigma_{i+1}|}{|[\bm\Lambda^\prime]_{\mathcal Q^\prime}|}
  {|(\bm\Gamma_i+\alpha^{\prime2}\bm\Gamma_i[\bm\Lambda^\prime]_{\mathcal Q^\prime}^{-1}
  \bm\Sigma_{i+1}[\bm\Lambda^\prime]_{\mathcal Q^\prime}^{-1}\bm\Gamma_i)|}
  \frac{ \mathcal{CN}\left(\mathbf w_{i+1},\bm\mu_{i+1},\bm\Sigma_{i+1}\right)}{\mathcal{CN}\left(\bm\nu_i,\mathbf 0,(\bm\Gamma_i+\alpha^{\prime2}\bm\Gamma_i[\bm\Lambda^\prime]_{\mathcal Q^\prime}^{-1}
  \bm\Sigma_{i+1}[\bm\Lambda^\prime]_{\mathcal Q^\prime}^{-1}\bm\Gamma_i)^{-1}\right)},
   \end{align}
where
  \begin{align}
  \boldsymbol\Sigma_{i+1}=&\left([\bm\Lambda^\prime]_{\mathcal Q^\prime}^{-1}-\alpha^{\prime2}[\bm\Lambda^\prime]_{\mathcal Q^\prime}^{-1}\bm\Gamma_i
   [\bm\Lambda^\prime]_{\mathcal Q^\prime}^{-1}\right)^{-1},\\
   \bm\mu_{i+1}=&\alpha\bm\Sigma_{i+1}[\bm\Lambda^\prime]_{\mathcal Q^\prime}^{-1}\bm\Gamma_i\left(\left(\frac{\sigma_n^{2\prime}}{\sigma_p^2} \mathbf I_{|\mathcal Q^\prime|}\right)^{-1} \tilde{\mathbf y}_{i}^\prime + \mathbf \Sigma_{i}^{-1} \boldsymbol \mu_{i}\right).
  \end{align}

Notice that the equations \eqref{eq:cscn_prod_0} and \eqref{Quotient} in the Appendix, and
the following properties are utilized in the above derivations.
\begin{align}
  \mathcal{CN}\left(\mathbf 0;-\alpha^{-1} [\mathbf \Lambda^{\prime}]_{\mathcal Q^{\prime}}\bm\nu_i,\alpha^{-2} [\mathbf \Lambda^{\prime}]_{\mathcal Q^{\prime}}\bm\Gamma_i^{-1} [\mathbf \Lambda^{\prime}]_{\mathcal Q^{\prime}}\right)=
  \frac{|\bm\Gamma_{i}^{-1}|}{|\alpha^{-2} [\mathbf \Lambda^{\prime}]_{\mathcal Q^{\prime}}\bm\Gamma_i^{-1} [\mathbf \Lambda^{\prime}]_{\mathcal Q^{\prime}}|}
  \mathcal{CN}\left(\bm\nu_i;\mathbf 0,\bm\Gamma^{-1}\right),
  \end{align}
  \begin{align}
  &\mathcal{CN}(\mathbf 0;\alpha\bm\Sigma_{i+1}[\bm\Lambda^\prime]_{\mathcal Q^\prime}^{-1}\bm\Gamma_i\bm\nu_i,\bm\Sigma_{i+1})=\frac{1}{|\pi\bm\Sigma_{i+1}|}
  \exp\left(-\alpha^2\bm\nu_i^H\bm\Gamma_i[\bm\Lambda^\prime]_{\mathcal Q^\prime}^{-1}\bm\Sigma_{i+1}[\bm\Lambda^\prime]_{\mathcal Q^\prime}^{-1}
  \bm\Gamma_i\bm\nu_i\right)\notag\\
  &=\frac{|\alpha^{-2}\bm\Gamma_i^{-1}
[\bm\Lambda^\prime]_{\mathcal Q^\prime}\bm\Sigma_{i+1}^{-1}
[\bm\Lambda^\prime]_{\mathcal Q^\prime}\bm\Gamma_i^{-1}|}{|\bm\Sigma_{i+1}|}
\mathcal{CN}(\bm\nu_i;\mathbf 0,\alpha^{-2}\bm\Gamma_i^{-1}
[\bm\Lambda^\prime]_{\mathcal Q^\prime}\bm\Sigma_{i+1}^{-1}
[\bm\Lambda^\prime]_{\mathcal Q^\prime}\bm\Gamma_i^{-1}),
  \end{align}

{So}, $ \Omega_{f_{A,i+1} \to \mathbf w_{i+1}}$ can be reexpressed from \eqref{eq:Omega_fa_w_0} and \eqref{simplify} as:
\begin{align}\label{beta i+1 to alpha i+1}
  \Omega_{f_{A,i+1} \to \mathbf w_{i+1}} =
  &\omega_{i}\alpha^{\prime-2|\mathcal Q^\prime|}\mathcal{CN}\left(0;\tilde{\mathbf y}_{i}^\prime,
  \frac{\sigma_n^{2\prime}}{\sigma_p^2} \mathbf I_{|\mathcal Q^\prime|}\right)
  \mathcal{CN}\left(\mathbf 0;\bm{\mu}_i,\bm{\Sigma}_i\right) \frac{\mathcal{CN}\left(0;\alpha^{\prime-1}\mathbf{w}_{i+1},
  \alpha^{\prime-2}[\mathbf \Lambda^{\prime}]_{\mathcal Q^{\prime}}\right)}
  {\mathcal{CN}\left(\mathbf 0;\bm\Gamma_i
  \left(\alpha^\prime [\mathbf \Lambda^{\prime}]_{\mathcal Q^{\prime}}^{-1}\mathbf{w}_{i+1}
  +\bm\nu_i\right), \bm\Gamma_i\right)} \notag \\
  %%%%%%%%%%%%%%%%%%%%%%%%%%%%%%%%%%%%%%%%%%
  =& \omega_{i}\alpha^{\prime-2|\mathcal Q^\prime|}\mathcal{CN}\left(0;\tilde{\mathbf y}_{i}^\prime,
  \frac{\sigma_n^{2\prime}}{\sigma_p^2} \mathbf I_{|\mathcal Q^\prime|}\right)
  \mathcal{CN}\left(\mathbf 0;\bm{\mu}_i,\bm{\Sigma}_i\right) \notag \\
  %%%%%%%%%%%%%%%%%%%%%%%%%%%%%%%%%%%%%%%%
  &\kern 10pt \times \frac{\alpha^{\prime2|\mathcal Q^\prime|}|\bm\Gamma_i||\bm\Sigma_{i+1}| \cdot {|(\bm\Gamma_i+\alpha^{\prime2}\bm\Gamma_i[\bm\Lambda^\prime]_{\mathcal Q^\prime}^{-1}
  \bm\Sigma_{i+1}[\bm\Lambda^\prime]_{\mathcal Q^\prime}^{-1}\bm\Gamma_i)|}}{|[\bm\Lambda^\prime]_{\mathcal Q^\prime}|
  \mathcal{CN}\left(\bm\nu_i,\mathbf 0,(\bm\Gamma_i+\alpha^{\prime2}\bm\Gamma_i[\bm\Lambda^\prime]_{\mathcal Q^\prime}^{-1}
  \bm\Sigma_{i+1}[\bm\Lambda^\prime]_{\mathcal Q^\prime}^{-1}\bm\Gamma_i)^{-1}\right)}
  \mathcal{CN}\left(\mathbf w_{i+1},\bm\mu_{i+1},\bm\Sigma_{i+1}\right)\notag \\
  %%%%%%%%%%%%%%%%%%%%%%%%%%%%%%%%%%%%%%%%%%
  =& \omega_{i+1} \mathcal {CN} \left(\mathbf w_{i+1}; \boldsymbol \mu_{i+1}, \mathbf \Sigma_{i+1} \right),
\end{align}
where
\begin{small}
\begin{align}\label{omega_i+1_2}
 \omega_{i+1} = \omega_{i} \mathcal{CN}\left(0;\tilde{\mathbf y}_{i}^\prime,
  \frac{\sigma_n^{2\prime}}{\sigma_p^2} \mathbf I_{|\mathcal Q^\prime|}\right)
  \mathcal{CN}\left(\mathbf 0;\bm{\mu}_i,\bm{\Sigma}_i\right)
  \frac{|\bm\Gamma_i||\bm\Sigma_{i+1}| \cdot
  {|(\bm\Gamma_i+\alpha^{\prime2}\bm\Gamma_i[\bm\Lambda^\prime]_{\mathcal Q^\prime}^{-1}
  \bm\Sigma_{i+1}[\bm\Lambda^\prime]_{\mathcal Q^\prime}^{-1}\bm\Gamma_i)|}}{|[\bm\Lambda^\prime]_{\mathcal Q^\prime}|
  \mathcal{CN}\left(\bm\nu_i,\mathbf 0,(\bm\Gamma_i+\alpha^{\prime2}\bm\Gamma_i[\bm\Lambda^\prime]_{\mathcal Q^\prime}^{-1}
  \bm\Sigma_{i+1}[\bm\Lambda^\prime]_{\mathcal Q^\prime}^{-1}\bm\Gamma_i)^{-1}\right)}.
\end{align}
\end{small}

%It can be seen from \figurename{ \ref{factor graph}} that this update rule holds for $1 \leq i\leq m-1 $.
%in every step $i$, the parameters
%$\omega_{i+1}$, $\boldsymbol \mu_{i+1}$, and $\mathbf \Sigma_{i+1} $ in $ \Omega_{f_{A,i+1} \to \mathbf w_{i+1}}$
%are related and only related to those in $ \Omega_{f_{A,i} \to \mathbf w_{i}}$.
%In addition, it is \textcolor{red}{checked} that $\Omega_{f_{A,1} \to \mathbf w_{1} } = \omega_1 \mathcal{CN} \left( \mathbf w_1; \mathbf \mu_1, \mathbf \Sigma_1 \right) $.

\end{lemma}

With Lemma \ref{lemma1}, we will finally obtain the message
$\Omega_{f_{A,m} \to \mathbf w_m} = \omega_{m} \mathcal {CN}(\mathbf w_m; \boldsymbol \mu_{m}, \mathbf \Sigma_{m})$.
Then the equation \eqref{y marginalization PDF} can be rewritten as:
\begin{align}\label{y|theta}
  p(\tilde{\mathbf y}^{\prime} (m)  | {\boldsymbol \vartheta}) &= \int_{\mathbf w_m} \Omega_{f_{A,m} \to \mathbf w_m} \Omega_{f_{B,m} \to \mathbf w_m} d \mathbf w_m \notag \\
  %%%%%%%%%%%%%%%%%%%%%%%%%%%%%%%%%%%%%%%%%%%%%%%
  &= \int_{\mathbf w_m}
  \mathcal {CN} \left(\tilde{\mathbf y}_{i}^\prime;  \mathbf w_m, \frac{\sigma_n^{2\prime}}{\sigma_p^2} \mathbf I_{|\mathcal Q^\prime|} \right)
  \omega_m  \mathcal {CN} \left(\mathbf w_m; \boldsymbol \mu_{m}, \mathbf \Sigma_{m} \right) d \mathbf w_m \notag \\
  %%%%%%%%%%%%%%%%%%%%%%%%%%%%%%%%%%%%%%%%%%%%%%
%  &= { \omega_m \int_{\mathbf w_m}
%  \mathcal {CN} \left(\mathbf w_m; \tilde{\mathbf y}_{m}^\prime, \frac{\sigma_n^{2\prime}}{\sigma_p^2} \mathbf I_{|\mathcal Q^\prime|} \right)
%  \mathcal {CN} \left(\mathbf w_m; \boldsymbol \mu_{m}, \mathbf \Sigma_{m} \right) d \mathbf w_m } \notag \\
  %%%%%%%%%%%%%%%%%%%%%%%%%%%%%%%%%%%%%%%%%%%%%%
  &= \omega_m \frac{ \mathcal {CN} \left(\mathbf 0; \tilde{\mathbf y}_{m}^\prime, \frac{\sigma_n^{2\prime}}{\sigma_p^2} \mathbf I_{|\mathcal Q^\prime|} \right)
  \mathcal {CN} \left(\mathbf 0; \boldsymbol \mu_{m}, \mathbf \Sigma_{m} \right) }{\mathcal{CN}\left(\mathbf 0; \mathbf G_{m}, \mathbf \Delta_{m} \right) },
  %&= \omega_m \frac{ | \mathbf \Delta_{m} | }{ | \mathbf \Sigma_{m} | }
%   \mathcal {CN} \left(\tilde{\mathbf y}_{m}^\prime; \mathbf 0_{N_t \times 1}, \frac{\sigma_n^{2\prime}}{\sigma_p^2} \mathbf I_{|\mathcal Q^\prime|} \right)
%    \times \text{exp} (\mathbf G_{m}^H \mathbf \Delta_{m}^{-1} \mathbf G_{m} - \boldsymbol \mu_{m}^H \mathbf \Sigma_{m}^{-1} \boldsymbol \mu_{m} )
\end{align}
where
\begin{align}\label{delta G m}
  \mathbf \Delta_{m}^{-1} = \left(\frac{\sigma_n^{2\prime}}{\sigma_p^2} \mathbf I_{|\mathcal Q^\prime|} \right)^{-1} + \mathbf \Sigma_{m}^{-1} ,
  \kern 30pt
  \mathbf G_{m} = \mathbf \Delta_{m}
  \left( (\frac{\sigma_n^{2\prime}}{\sigma_p^2} \mathbf I_{|\mathcal Q^\prime|})^{-1} \tilde{\mathbf y}_{m}^\prime + \mathbf \Sigma_{m}^{-1} \boldsymbol \mu_{m} \right).
\end{align}

Hence, using the adopted sum-product and factor graph algorithm,
the likelihood function $p(\tilde{\mathbf y}^{\prime} (m)  | {\boldsymbol \vartheta}) $
can be obtained according to \eqref{y|theta},
where all the parameters defined before can be obtained according to the above recursion processes.

\subsubsection{\textbf{MCMC computation}}
As the two posterior effective noise statistics
$\mathbb E_{\boldsymbol \vartheta} \left\{\frac{\sigma_n^{2\prime}}{\sigma_p^2} \mathbf I_{|\mathcal Q^\prime|} | \tilde{\mathbf y}^{\prime} (m) \right\}$
and $\mathbb E_{\boldsymbol \vartheta} \left\{[\mathbf \Lambda^{\prime}]_{\mathcal Q^{\prime}} | \tilde{\mathbf y}^{\prime} (m) \right\} $ are unknown,
we employ the Metropolis Hastings MCMC algorithm to estimate them.
This algorithm is used to the case where the proposal distribution is no longer a symmetric function of its arguments \cite{PR}.
At the $j $-th iteration, the last accepted MCMC sample in the sequence of samples be ${\boldsymbol \vartheta}^{(j)}$ generated.
A candidate MCMC sample $\tilde {\boldsymbol \vartheta}$ will be generated according to a proposed distribution $p(\tilde {\boldsymbol \vartheta} | {\boldsymbol \vartheta}^{(j)})$.
As the specific choice of proposal distribution can have a prominent effect on the performance of the algorithm,
we choose a Gaussian distribution centred on the current state $\boldsymbol \vartheta^{(j)}$.
The candidate MCMC sample $\tilde {\boldsymbol \vartheta} $ will be either accepted or rejected according to an acceptance ratio $r$ defined as
\begin{align}\label{ratio r}
  r &= \text{min} \left\{1, \frac{p({\boldsymbol \vartheta}^{(j)}|\tilde {\boldsymbol \vartheta})
  p(\tilde{\mathbf y}^{\prime} (m) | \tilde {\boldsymbol \vartheta}) p(\tilde {\boldsymbol \vartheta})}
  {p(\tilde {\boldsymbol \vartheta} | {\boldsymbol \vartheta}^{(j)})
  p(\tilde{\mathbf y}^{\prime} (m) | {\boldsymbol \vartheta}^{(j)}) p({\boldsymbol \vartheta}^{(j)}) } \right\} \notag \\
  %%%%%%%%%%%%%%%%%%%%%%%%%%%%%%%%%%%%%%%%%%%%%%%
  &= \text{min} \left\{1, \frac{p(\tilde{\mathbf y}^{\prime} (m) | \tilde {\boldsymbol \vartheta}) p(\tilde {\boldsymbol \vartheta})}
  {p(\tilde{\mathbf y}^{\prime} (m) | {\boldsymbol \vartheta}^{(j)}) p({\boldsymbol \vartheta}^{(j)}) } \right\},
\end{align}
where the second formula is used when the proposal distribution is symmetric,
i.e., $p(\tilde {\boldsymbol \vartheta} | {\boldsymbol \vartheta}^{(j)}) = p({\boldsymbol \vartheta}^{(j)}|\tilde {\boldsymbol \vartheta})  $.
The $(j+1) $-th MCMC sample is
\begin{align}\label{theta j+1}
 {\boldsymbol \vartheta}^{(j+1)} = \left\{
 \begin{aligned}
 &\tilde {\boldsymbol \vartheta} &\text{with probability}\  r, \\
 &{\boldsymbol \vartheta}^{(j)} &\text{otherwise}.
 \end{aligned}
 \right.
\end{align}

We can iterate the process in (\ref{ratio r}), (\ref{theta j+1}),
and achieve a sequence of MCMC samples.
The positivity of the proposal distribution $p(\tilde {\boldsymbol \vartheta}|{\boldsymbol \vartheta}^{(j)}) $
for any ${\boldsymbol \vartheta}^{(j)} $ is a sufficient condition for an ergodic Markov
chain of MCMC samples, whose steady-state distribution is the target distribution $ p({\boldsymbol \vartheta} | \tilde{\mathbf y}^{\prime} (m))$.
After generating enough MCMC samples, the posterior effective noise statistics can be approximated
by computing the sample mean of the accepted MCMC samples.
%For the simulations in this paper, we use a Gaussian distribution as the proposal distribution in the MCMC step.

\begin{figure}[!t]
	\centering
	\includegraphics[width=160mm]{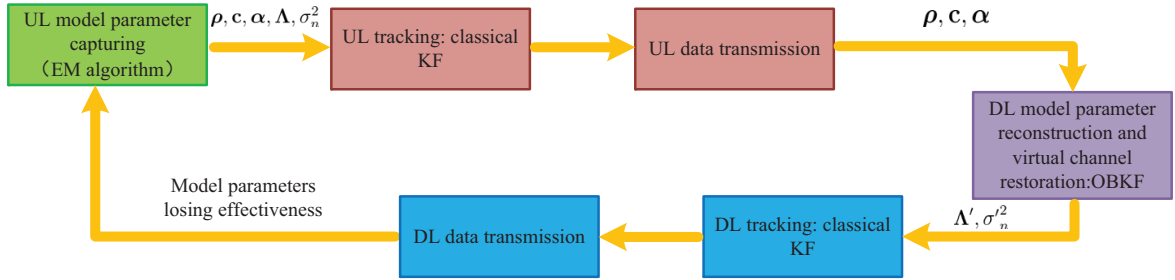}
	\caption{ The algorithm block diagram of the proposed scheme. }
	\label{fig:algorithm_block}
\end{figure}

%Since $\alpha_k^\prime$ is well estimated from UL process and used in the KF tracking,
%the estimation of KF is more accurate than the estimation obtained directly from measurement equation \eqref{eq:AR_downlink} by LS estimator,
%although there is no deviation in $\big[ {{\mathbf \Lambda}}_{k}^{\prime}\big]_{\mathcal Q_k^{\prime}}$.
The steps of the whole procedure for the DL channel reconstruction and restoration are summarized in Algorithm \ref{alg:learning_and_tracking}.
And In order to describe the relationship among different parts of our proposed scheme intuitively,
the overall algorithm block diagram of the proposed scheme are illustrated in \figurename{ \ref{fig:algorithm_block}}.

\begin{algorithm}
	\caption{UL Learning Aided DL Reconstruction and Restoration}
	\label{alg:learning_and_tracking}
	\renewcommand{\arraystretch}{0.5}
	\begin{algorithmic}[1]
        \STATE {\bf input:} $ \mathbf y,  p(\theta), \tilde{\mathbf y}^{\prime} (m) $.
%		\STATE {\bf output:} $[\widehat {\tilde{\mathbf g}}_{k,{m}}]_{\mathcal Q_k^{\prime}}^{\theta} $
        \STATE {\bf initialize:} $\boldsymbol{\hat \alpha}^{(0)}, \mathbf{\hat \Lambda}^{(0)}, \mathbf{\hat c}^{(0)}, \boldsymbol{\hat \rho}^{(0)}, {\hat {\sigma_n^2}}^{(0)}, \quad \theta^{(0)} \sim \pi(\theta) $.
        \FOR {$l = 0, 1, \ldots, l$}
        \STATE $\mathbf{\hat c}^{(l)} \gets $ \bf{Algorithm \ref{alg:seraching_c}}, $[\hat {\boldsymbol \rho}_k^{(l)}]_j \gets \eqref{update_rho} $ \bf{and the constraint}  $[{\boldsymbol \rho}_k]_j \in [-\frac{1}{2}, \frac{1}{2}]  $.
        \STATE $\boldsymbol{\hat \alpha}^{(l)} \gets \eqref{update_alpha}, \quad [\hat{\boldsymbol\Lambda}_k]_{j,j}^{(l)} \gets \eqref{update_Lambda}, \quad {\hat {\sigma_n^2}}^{(l)} \gets \eqref{update_sigma}$.
%        \STATE $[\hat{\boldsymbol\Lambda}_k]_{j,j}^{(l)} \gets \eqref{update_Lambda} $
%        \STATE ${\hat {\sigma_n^2}}^{(l)} \gets \eqref{update_sigma} $
       % \STATE $\mathbf{\hat c}^{(l)} \gets \textbf{Searching by Algorithm 1 in \cite{Ma_J_SBL_Time_varing}} $
%        \STATE $[\hat {\boldsymbol \rho}_k^{(l)}]_j \gets \eqref{update_rho} \textbf{and the constraint} [{\boldsymbol \rho}_k]_j \in [-\frac{1}{2}, \frac{1}{2}] $
        \ENDFOR
        \STATE $\alpha_k^\prime \gets \eqref{DL_alpha}$, $\mathcal Q_k^{\prime}, \boldsymbol \rho_k^{\prime} \gets \eqref{angle reciprocity}, \eqref{p' and rho}$.
		\STATE $ [\widehat {\tilde{\mathbf g}}_{0}]_{\mathcal Q^{\prime}}^{\boldsymbol \vartheta} \gets \mathbb E\left\{[{\tilde{\mathbf g}}_{m}]_{\mathcal Q^{\prime}}^{\boldsymbol \vartheta}\right\}, \quad
    \mathbb E_{\boldsymbol \vartheta} [\mathbf P_0^{\boldsymbol \vartheta} | \tilde{\mathbf y}^{\prime} (0)] \gets \mathrm{cov} \left[ [{\tilde{\mathbf g}}_{m}]_{\mathcal Q^{\prime}}^{\boldsymbol \vartheta} \right]$.
        \STATE $ \mathbb E_{\boldsymbol \vartheta} \left\{[\mathbf \Lambda^{\prime}]_{\mathcal Q^{\prime}} | \tilde{\mathbf y}^{\prime} (0) \right\} \gets \mathbb E_{\boldsymbol \vartheta} \left\{[\mathbf \Lambda^{\prime}]_{\mathcal Q^{\prime}} \right\},
        \quad \mathbb E_{\boldsymbol \vartheta}\left\{\frac{\sigma_n^{2\prime}}{\sigma_p^2} \mathbf I_{|\mathcal Q^\prime|} | \tilde{\mathbf y}^{\prime} (0) \right\} \gets \mathbb E_{\boldsymbol \vartheta} \left\{ \frac{\sigma_n^{2\prime}}{\sigma_p^2} \mathbf I_{|\mathcal Q^\prime|} \right\} $.
		
        \FOR {$m=1,2, \ldots $}
		\STATE $\tilde{\mathbf z}_{m}^{\boldsymbol \vartheta} \gets \tilde{\mathbf y}_{m}^{\prime {\boldsymbol \vartheta}}
   - [\widehat {\tilde{\mathbf g}}_{m}]_{\mathcal Q^{\prime}}^{\boldsymbol \vartheta}$.
        \STATE $\mathbf K_{m}^{\Theta} \gets \mathbb E_{\boldsymbol \vartheta} \left\{\mathbf P_{m}^{\boldsymbol \vartheta} | \tilde{\mathbf y}^{\prime} (m-1) \right\}
   \mathbb E_{\boldsymbol \vartheta}^{-1} \left\{ \mathbf P_{m}^{\boldsymbol \vartheta} + \frac{\sigma_n^{2\prime}}{\sigma_p^2} \mathbf I_{|\mathcal Q^\prime|} | \tilde{\mathbf y}^{\prime} (m-1) \right\} $.
        \FOR {$ i = 1, 2, \ldots, \text {num\_iterations}$}
		\STATE $\tilde {\boldsymbol \vartheta} \sim p( {\boldsymbol \vartheta} | {\boldsymbol \vartheta}^{(i-1)})$.
        \FOR {$ j = 2, 3, \ldots,  m-1$}
        \STATE $\bm\nu_j \gets \left(\frac{\sigma_n^{2\prime}}{\sigma_p^2} \mathbf I_{|\mathcal Q^\prime|}\right)^{-1} \tilde{\mathbf y}_{j}^\prime +
        \mathbf \Sigma_{j}^{-1} \boldsymbol \mu_{j}, \kern 10pt
        \mathbf \Gamma_{j} \gets \left(\alpha^{\prime 2} [\mathbf \Lambda^{\prime}]_{\mathcal Q^{\prime}}^{-1} + (\frac{\sigma_n^{2\prime}}{\sigma_p^2} \mathbf I_{|\mathcal Q^\prime|})^{-1} + \mathbf \Sigma_{j}^{-1} \right)^{-1} $.
        \STATE $\boldsymbol\Sigma_{j+1} \gets \left([\bm\Lambda^\prime]_{\mathcal Q^\prime}^{-1}-\alpha^{\prime2}[\bm\Lambda^\prime]_{\mathcal Q^\prime}^{-1}\bm\Gamma_j
        [\bm\Lambda^\prime]_{\mathcal Q^\prime}^{-1}\right)^{-1} $.
        \STATE $ \bm\mu_{j+1} \gets \alpha^\prime \bm\Sigma_{j+1}[\bm\Lambda^\prime]_{\mathcal Q^\prime}^{-1}\bm\Gamma_j
        \left(\left(\frac{\sigma_n^{2\prime}}{\sigma_p^2} \mathbf I_{|\mathcal Q^\prime|}\right)^{-1} \tilde{\mathbf y}_{j}^\prime + \mathbf \Sigma_{j}^{-1} \boldsymbol \mu_{j}\right)  $.
        \STATE $\omega_{j+1} \gets \text{using} \eqref{omega_i+1} $.
		\ENDFOR
        \STATE $\mathbf \Delta_{m}^{-1} \gets \left(\frac{\sigma_n^{2\prime}}{\sigma_p^2} \mathbf I_{|\mathcal Q^\prime|} \right)^{-1} + \mathbf \Sigma_{m}^{-1} ,
        \quad \mathbf G_{m} \gets \mathbf \Delta_{m} \left( (\frac{\sigma_n^{2\prime}}{\sigma_p^2} \mathbf I_{|\mathcal Q^\prime|})^{-1} \tilde{\mathbf y}_{m}^\prime + \mathbf \Sigma_{m}^{-1} \boldsymbol \mu_{m} \right) $.
        \STATE $p(\tilde{\mathbf y}^{\prime} (m)  | {\boldsymbol \vartheta}) \gets \text{using} \eqref{y|theta},
        \quad p(\tilde {\boldsymbol \vartheta} | \tilde{\mathbf y}^{\prime} (m)) \gets p(\tilde{\mathbf y}^{\prime} (m)  | {\boldsymbol \vartheta}) $.
        \STATE $r \gets \text{min} \{1, \frac{p(\tilde{\mathbf y}^{\prime} (m) | \tilde {\boldsymbol \vartheta}) p(\tilde {\boldsymbol \vartheta})}
  {p(\tilde{\mathbf y}^{\prime} (m) | {\boldsymbol \vartheta}^{(i-1)}) p({\boldsymbol \vartheta}^{(i-1)}) } \} , \quad \zeta \sim \text{unif}(0,1)$.
        \IF {$\zeta < r $}
        \STATE ${\boldsymbol \vartheta}^{(i)} \gets {\tilde {\boldsymbol \vartheta}} $.
        \ELSE
        \STATE ${\boldsymbol \vartheta}^{(i)} \gets {\boldsymbol \vartheta}^{(i-1)} $.
        \ENDIF
		\ENDFOR
        \STATE $\mathbb E_{\boldsymbol \vartheta} \left\{\frac{\sigma_n^{2\prime}}{\sigma_p^2} \mathbf I_{|\mathcal Q^\prime|} | \tilde{\mathbf y}^{\prime} (m) \right\},
        \mathbb E_{\boldsymbol \vartheta} \left\{ [\mathbf \Lambda^{\prime}]_{\mathcal Q^{\prime}} | \tilde{\mathbf y}^{\prime} (m)\right\}
        \gets \{{\boldsymbol \vartheta}^{(1)}, {\boldsymbol \vartheta}^{(2)}, \ldots {\boldsymbol \vartheta}^{(m)} \} $.
        \STATE $\mathbb E_{\boldsymbol \vartheta} \left\{\mathbf P_{m+1}^{\boldsymbol \vartheta} | \tilde{\mathbf y}^{\prime} (m) \right\}
        \gets \alpha^{\prime2} (\mathbf I - \mathbf K_{m}^{\Theta *} \mathbf )
   \mathbb E_{\boldsymbol \vartheta} \left\{\mathbf P_m^{\boldsymbol \vartheta} | \tilde{\mathbf y}^{\prime} (m)\right\}
   + \mathbb E_{\boldsymbol \vartheta}  \Big\{ [\mathbf \Lambda^{\prime}]_{\mathcal Q^{\prime}} | \tilde{\mathbf y}^{\prime} (m)  \Big\} $.
        \STATE $[\widehat {\tilde{\mathbf g}}_{{m+1}}]_{\mathcal Q^{\prime}}^{\boldsymbol \vartheta} \gets
        \alpha^\prime [\widehat {\tilde{\mathbf g}}_{m}]_{\mathcal Q^{\prime}}^{\boldsymbol \vartheta}
   + \alpha^\prime \mathbf K_{m}^{\Theta} \tilde{\mathbf z}_{m}^{\boldsymbol \vartheta}$.
        \RETURN{ $[\widehat {\tilde{\mathbf g}}_{{m+1}}]_{\mathcal Q^{\prime}}^{\boldsymbol \vartheta} $. }
        \ENDFOR		
	\end{algorithmic}
\end{algorithm}

\section{Simulations Results}
In this section, we will evaluate the performance of our proposed tracking scheme through numerical simulation.
We consider a massive MIMO network where the BS is equipped with $N_t = 128 $ antennas.
$K = 32 $ is the number of users, while they are divided into $8$ groups.
We take the first group as an example to show the perfect performance.
The simulation parameters are summarized in {TABLE 1}.
\begin{table}[!t]
	\centering
	\renewcommand{\arraystretch}{1.0}
	\caption{Simulation Parameters}
	\label{table}
	\begin{tabular}{c|c}
		\hline
		\hline
		Number of BS antennas $N_t$ &128\\
		\hline
		Number of users per group $\tau$&4 \\	\hline
		angle spread range&$[-49^\circ -43^\circ], [-26^\circ -20^\circ],$
		\\  &$[20^\circ 26^\circ],[43^\circ 49^\circ]$\\
		\hline	
		Length	of  training sequences $L_t$ & 4\\
		\hline
		Channel	coherence interval $L_c$& 160\\
		\hline
		Symbol period & 1 us\\
		\hline
		Carrier frequency &  2 GHz\\
		\hline
		BS antenna space &  $\lambda/2$\\
		\hline
		\hline
		
	\end{tabular}	
\end{table}

Correspondingly, the preamble is also divided into $8$ segments.
Only $4$ users in the same group are active tt each segment, so there is no inter-group interference.
As a result, the training of length 4 is sufficient.
During the virtual channel tracking stage,
the $K$ users are regrouped such that the spatial signatures of the users in the same group do not overlap.
The signal-to-noise ratio (SNR) is defined as SNR$= \sigma_p^2/\sigma_n^2$.
The performance metrics are taken as the
average  MSEs of  the model parameters $\alpha$, $\mathbf c$ and $\mathbf \Lambda$  and that of the virtual channel $\tilde{\mathbf g}$ and $\tilde{\mathbf h}$ i.e.,
\begin{align}
%\text{MSE}_{\alpha} =& \frac{1}{\tau} \sum_{i=1}^{\tau} \frac{|\hat{\alpha}_i-\alpha_i|^2}{|\alpha_i|^2}\\
\text{MSE}_{\mathbf x} =& \frac{1}{\tau} \sum_{i=1}^{\tau} \frac{\|\hat{\mathbf x}_i-\mathbf x_i\|^2}{\|\mathbf x_i\|^2}, {\mathbf x} = \alpha, \mathbf c, \mathbf \Lambda, \boldsymbol \rho, \sigma_n^2, \tilde{\mathbf h},\tilde{\mathbf g}.
%\text{MSE}_{\mathbf \Lambda} =& \frac{1}{\tau} \sum_{i=1}^{\tau} \frac{\|\text{diag}(\hat{\mathbf \Lambda}_i)-\text{diag}(\mathbf   \Lambda_i)\|^2}{\|\text{diag}(\mathbf \Lambda_i)\|^2}
\end{align}

\begin{figure}[!t]
	\centering
	\includegraphics[width=70mm]{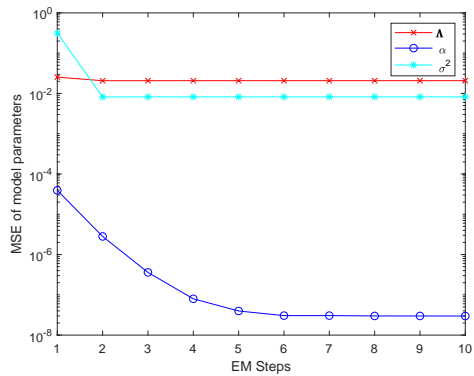}
	\caption{ The convergence  of the EM based UL model parameter learning algorithm with SNR = 30dB.}
	\label{fig:EM}
\end{figure}

We first investigate the  convergence of the UL EM process.
\figurename{ \ref{fig:EM}} shows the MSEs curves versus the number of iteration.
$M_u=15$ channel blocks are used to learn model parameters.
We can see from \figurename{ \ref{fig:EM}} that after 5 iterations, all the parameters have arrived at their steady states, which shows that the algorithm has a fast convergence speed.

%\begin{figure}[!t]
%	\centering
%	\includegraphics[width=85mm]{MSE_parameter_Vs_SNR_M50.eps}
%	\caption{The MSE performance of the UL model parameters learning versus SNR. }
%	\label{fig:model_SNR}
%\end{figure}

\begin{figure}[htbp]
  \centering
  \subfigure[]
  {\label{fig:model_SNR}
  \begin{minipage}{60mm}
  \centering
  \includegraphics[width=50mm]{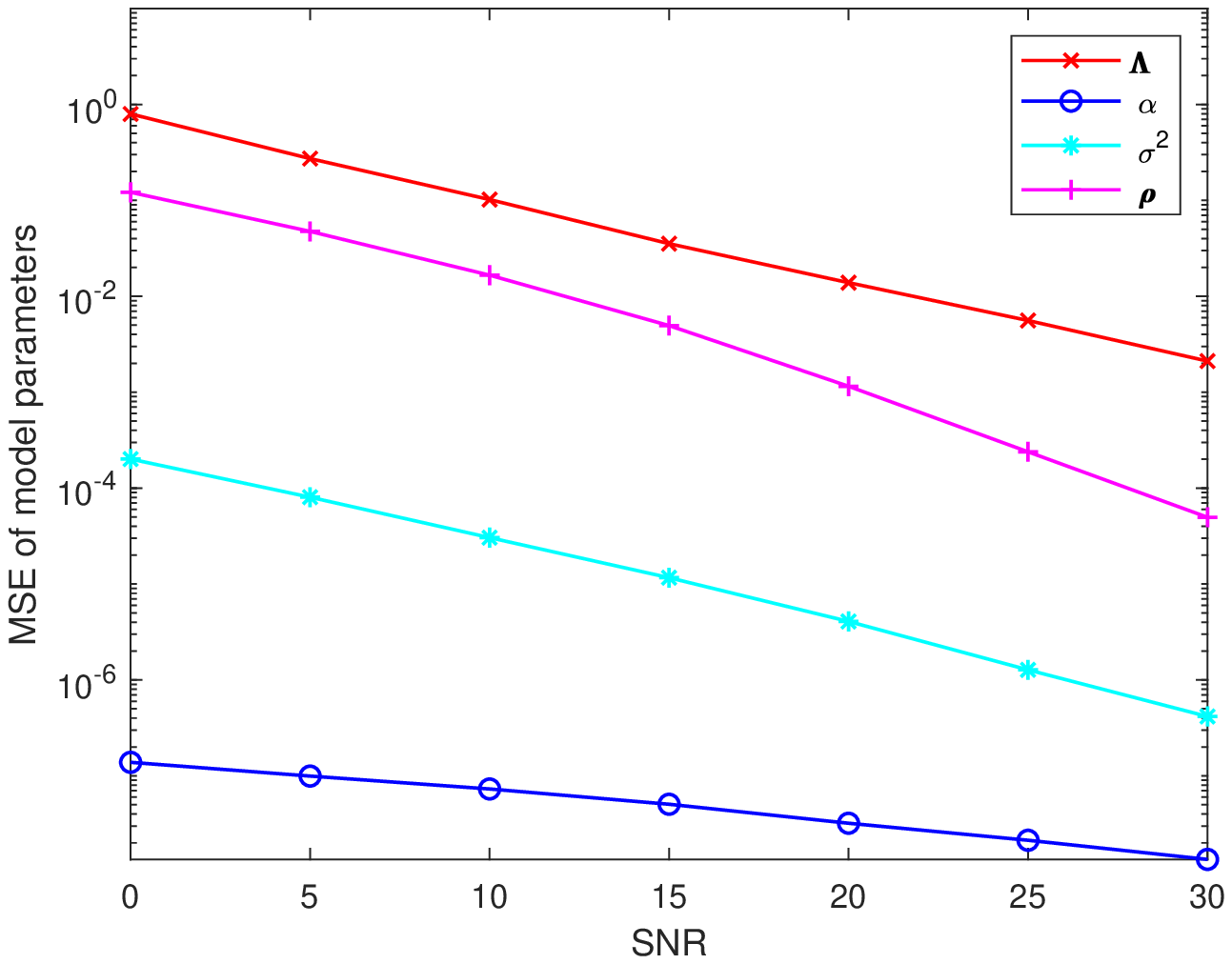}
  \end{minipage}
  }
  \subfigure[%An image expression for algorithm \ref{alg:seraching_c}.
  ]
  {\label{fig:bias_compare}
  \begin{minipage}{60mm}
  \centering
  \includegraphics[width=50mm]{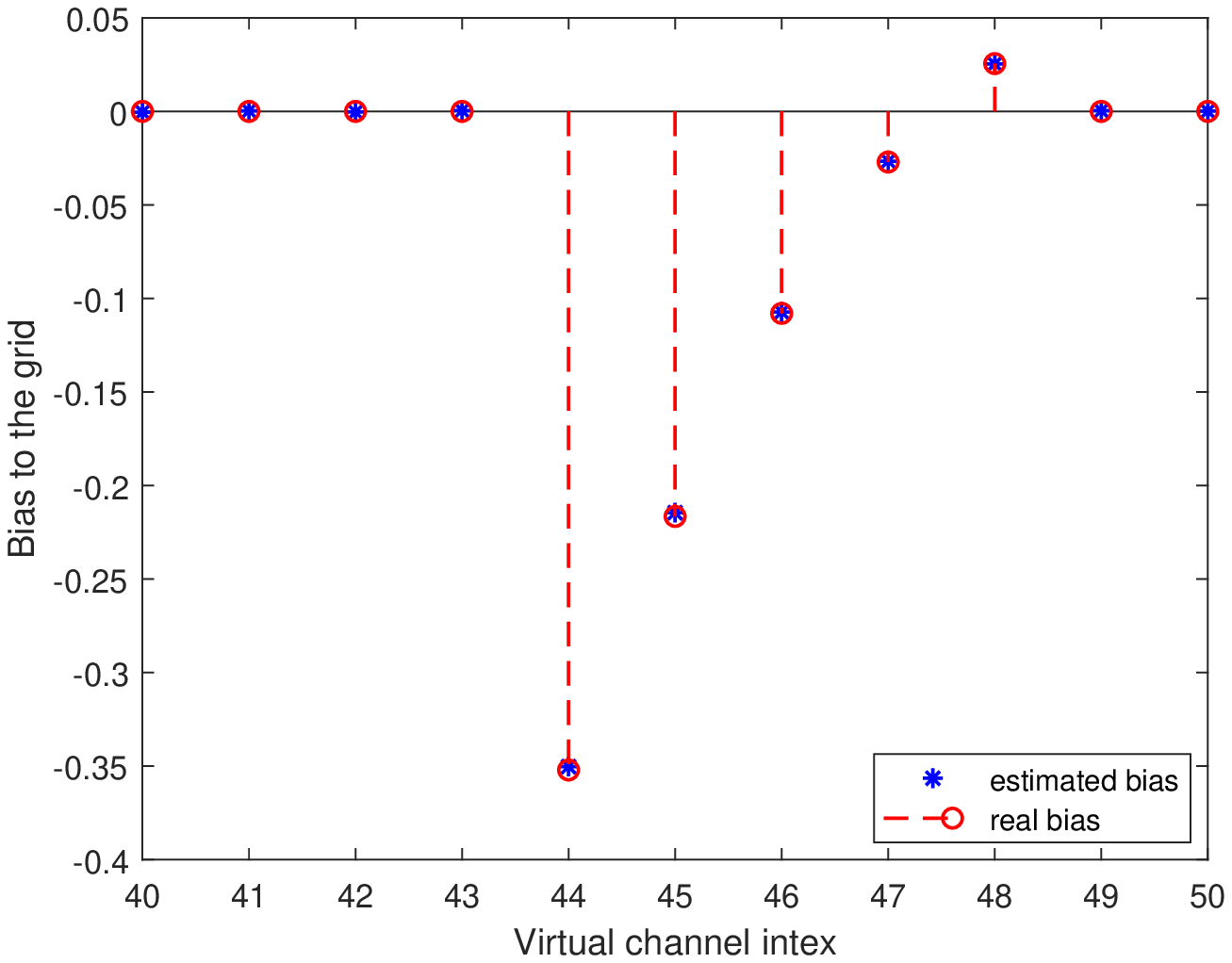}
  \end{minipage}
  }
  \caption{(a) The MSE performance of the UL model parameters learning versus SNR.
  (b) The off-grid bias and spatial signature performance with SNR = 20dB. }
\end{figure}

\figurename{ \ref{fig:model_SNR}} presents the MSE performance of the model parameters learning as a function of SNR, with EM algorithm running 5 iterations for each SNR case.
With the increase of the SNR, we can see that the MSE curves of all parameters decrease almost linearly.
Moreover, we show the performance of the estimation for the off-grid bias and spatial signature performance in \figurename{ \ref{fig:bias_compare}}, with SNR = 20dB.
They are also estimated very accurately.

After UL learning of all parameters and DL reconstruction of partial parameters, the next step is to track the DL channel by adopt OBKF, with the known parameters, meanwhile restore the unreconstructed parameters for later tracking. To decrease the computation complexity, we will adopt OBKF for a limited number of time-blocks, and then use classical KF to continue tracking the channel.

\begin{figure}[!t]
	\centering
	\includegraphics[width=70mm]{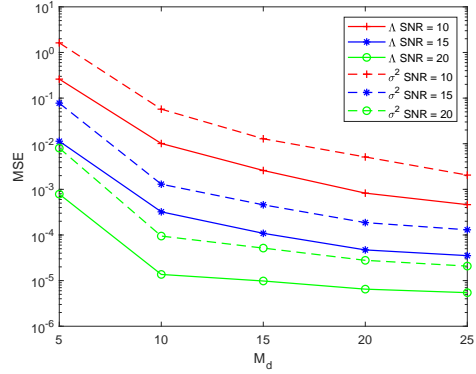}
	\caption{The MSE of the DL channel model parameters versus OBKF blocks $M_d$. }
	\label{fig:DLMSE_M_snrindex}
\end{figure}

\begin{figure}[!t]
	\centering
	\includegraphics[width=70mm]{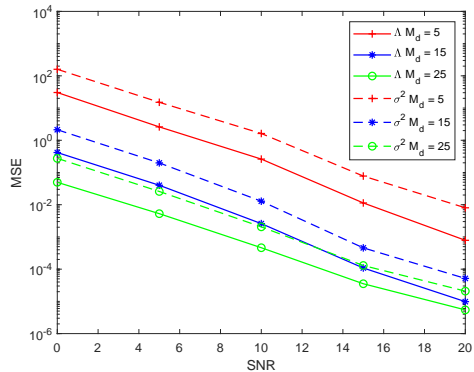}
	\caption{The MSE of the DL virtual channel model parameters versus SNR. }
	\label{fig:DLMSE_snr_M}
\end{figure}

First, we studies the MSE of the two unknown DL channel model parameters at the last OBKF time-block versus the number of time-block using OBKF, with different SNR, and the MSE versus SNR with different number of OBKF time-block.
In \figurename{ \ref{fig:DLMSE_M_snrindex}},
we can see that the MSE of the two unknown DL channel model parameters decreases in each SNR case, and almost arrive at their convergence point when OBKF time-block $ M_d=15$.
As SNR goes higher, the convergence point can be arrived when OBKF time-block $ M_d=10$.
We can also see that the curves linearly decrease with the increase of the SNR, in \figurename{ \ref{fig:DLMSE_snr_M}}.

We can find that the performance are better when SNR is higher.
we can explain the above phenomenon that OBKF is not only restoring the virtual channel, but also restoring the unknown parameters.
And after a scale of restoring time, the parameters will be very closed to the true one,
so the estimated virtual channel will have a good performance.

\begin{figure}[!t]
	\centering
	\includegraphics[width=70mm]{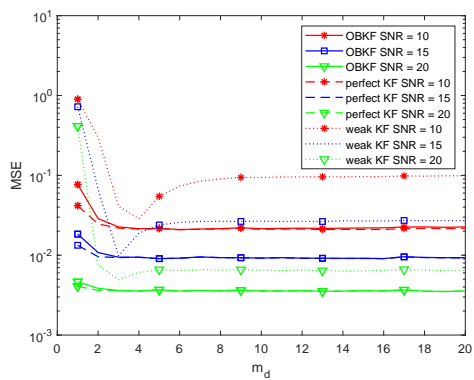}
	\caption{The MSE of the DL virtual channel for each tracking blocks $m_d$. }
	\label{fig:DLMSE_m}
\end{figure}

Then we studied the MSE of virtual channel for each time-block, including both the OBKF time-blocks and the later classical KF time-blocks, with different SNR, as shown in \figurename{ \ref{fig:DLMSE_m}}.
The figure also shows the performance of classical KF with perfect parameters as well as classical KF with weak parameters.
We set the number of OBKF time-block $M_d=10$, with which we can obtain almost the best performance.
From \figurename{ \ref{fig:DLMSE_m}} we can obtain that the MSE of tracked virtual channel decreases when OBKF runs.
We can see that the performance to be steady and is very close to the performance of classical KF with perfect parameters at $m_d = 6 $, which shows the accuracy of our method.

\begin{figure}[!t]
	\centering
	\includegraphics[width=70mm]{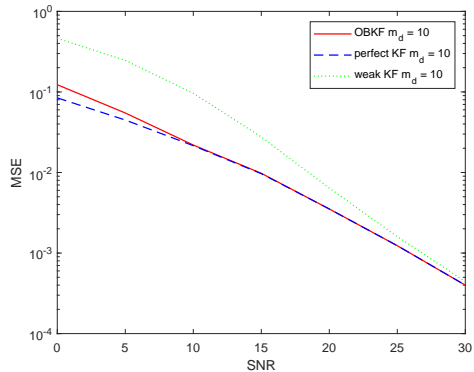}
	\caption{The MSE of the DL virtual channel versus SNR. }
	\label{fig:DLMSE_snr_m}
\end{figure}

To further illustrate the performance of the method,
\figurename{ \ref{fig:DLMSE_snr_m}} shows the relationship between MSE of virtual channel and SNR, together with classical KF of the above two situations.
from \figurename{ \ref{fig:DLMSE_snr_m}} we can see that the performance of KF with weak parameters is far away from the precisely one,
while our OBKF method has a wonderful performance.
Moreover, with the SNR increasing, the gap between our method and perfect KF decreases very fast. At SNR = 30, for example, the two performance is very nearly equal.
Notice that the gap between our method and weak KF is also decreasing.
This can be explained as follows.
At low SNR, our method obtains a huge gain by utilizing the correlation of time-varying channel.
But with SNR increasing, the performance is mostly decided on SNR,
meanwhile the effect of correlation is diminishing.

\begin{figure}[!t]
	\centering
	\includegraphics[width=70mm]{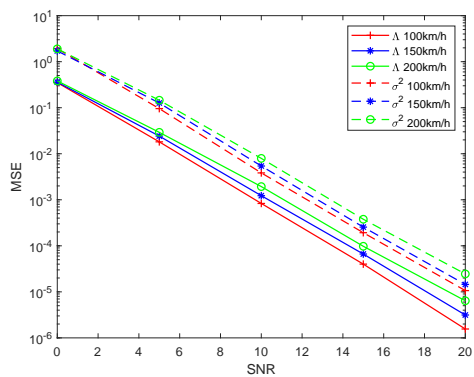}
	\caption{The MSE of the DL virtual channel model parameters for different velocity. }
	\label{fig:DLMSE_velocity}
\end{figure}

Furthermore, we show the MSE performance of the two unknown DL channel model parameters versus SNR for different velocity at the last OBKF time-block, while $M_d=15$.
In \figurename{ \ref{fig:DLMSE_velocity}}, we can find that the performance is better at slower velocity,
while at higher velocity the performance is only a little worse and is acceptable.

\section{Conclusion}
In this paper, we proposed a skillful scheme for the DL channel tracking.
First, with the help of VCR, a dynamic uplink (UL) massive MIMO channel model was built with the consideration of off-grid refinement.
Then, a coordinate-wise maximization based expectation maximization (EM) algorithm was adopted in the model parameters learning period.
Thanks to the angle reciprocity, with the knowledge of UL channel model parameters, we recovered some of the parameters of DL channel model.
After that, {as there remains some parameters which could not be perfectly inferred from the UL ones,}
we {resorted to} OBKF method to accurately track the DL channel.
During the method, factor-graph and Metropolis Hastings MCMC were applied to track the expectation of posterior statistics.
Numerical results showed that our proposed scheme has not only a strong convergence, but also a very low estimation MSE.





%\bibliographystyle{IEEEtran}
\balance
%\bibliographystyle{IEEEtran}
%\bibliography{./bibtex/IEEEabrv,./bibtex/ref}

\section{Appendix\\ The product of the $N$-dimensional complex Gaussian PDF}

For the $N$-dimensional complex Gaussian distribution $p(\mathbf x)=\mathcal{CN}\left({\boldsymbol x;\boldsymbol\mu,\boldsymbol\Sigma}\right)$, we can obtain its canonical notation as
\begin{align}
p(\mathbf x)=&\frac{1}{\pi^N|\boldsymbol\Sigma|}\exp(-(\mathbf x-\boldsymbol\mu)^H\boldsymbol\Sigma^{-1}(\mathbf x-\boldsymbol\mu))\notag\\
=&\exp(-\ln\pi^N-\ln|\boldsymbol\Sigma|)\exp\left\{-\mathbf x^H\boldsymbol\Sigma^{-1}\mathbf x
-\boldsymbol\mu^H\boldsymbol\Sigma^{-1}\boldsymbol\mu
+2\Re\{\mathbf x^H\boldsymbol\Sigma^{-1}\boldsymbol\mu\}\right\}\notag\\
=&\exp({-N\ln\pi-\ln|\boldsymbol\Sigma|-\boldsymbol\mu^H\boldsymbol\Sigma^{-1}\boldsymbol\mu}
-\mathbf x^H\boldsymbol\Sigma^{-1}\mathbf x+2\Re\{\mathbf x^H\boldsymbol\Sigma^{-1}\boldsymbol\mu\}).
\end{align}

Then, for the PDFs $p_i(\mathbf x)=\mathcal{CN}\left({\boldsymbol x;\boldsymbol\mu_i,\boldsymbol\Sigma_i}\right)$,
$i=1,2,\ldots, L$, we can derive
\begin{align}
\prod_{i=1}^L p_i(\mathbf x)=&\prod_{i=1}^L
\mathcal{CN}\left({\boldsymbol x;\boldsymbol\mu_i,\boldsymbol\Sigma_i}\right)\notag\\
=&\exp\left(\sum_{i=1}^L\zeta_i-\mathbf x^H\left(\sum_{i=1}^L\boldsymbol\Sigma_i^{-1}\right)\mathbf x
+2\Re\left\{\mathbf x^H\left(\sum_{i=1}^L \boldsymbol\Sigma^{-1}_i\boldsymbol\mu_i\right)\right\}\right),
\end{align}
where the term $\zeta_i=-N\ln\pi-\ln|\boldsymbol\Sigma_i|-\boldsymbol\mu_i^H\boldsymbol\Sigma^{-1}_i\boldsymbol\mu_i$
is defined in the above equation. Before proceeding, let us define
$\boldsymbol{\bar\Sigma}_L=\left(\sum_{i=1}^L\boldsymbol\Sigma_i^{-1}\right)^{-1}$,
and $\boldsymbol{\bar\mu}_L=\boldsymbol{\bar\Sigma}_L\left(\sum\limits_{i=1}^L \boldsymbol\Sigma^{-1}_i\boldsymbol\mu_i\right)$. Hence, the above equation can be reexpressed as
\begin{align}
\prod_{i=1}^L p_i(\mathbf x)=&\exp\left(\sum_{i=1}^L\zeta_i-\mathbf x^H\boldsymbol{\bar\Sigma}_L^{-1}\mathbf x
+2\Re\left\{\mathbf x^H\boldsymbol{\bar\Sigma}_L^{-1}\boldsymbol{\bar\mu}_L\right\}\right)\notag\\
=&\exp\left(\sum_{i=1}^L\zeta_i-\bar\zeta_L+\bar \zeta_L-
\mathbf x^H\boldsymbol{\bar\Sigma}_L^{-1}\mathbf x
+2\Re\left\{\mathbf x^H\boldsymbol{\bar\Sigma}_L^{-1}\boldsymbol{\bar\mu}_L\right\}\right)\notag\\
=&\exp\left(\sum_{i=1}^L\zeta_i-\bar\zeta_L\right)
\mathcal{CN}\left(\mathbf x; \boldsymbol{\bar\mu}_L, \boldsymbol{\bar\Sigma}_L\right),
\label{eq:cscn_prod_0}
\end{align}
where $\bar\zeta_L=-N\ln\pi-\ln|\boldsymbol{\bar\Sigma}_L|-\boldsymbol{\bar\mu}_L^H
\boldsymbol{\bar\Sigma}_L^{-1}\boldsymbol{\bar\mu}_L$.

Specially, for  $L=2$,  it can  be obtained that
\begin{align}
\exp\left(\sum_{i=1}^2\zeta_i-\bar\zeta_2\right)
=\frac{\mathcal{CN}(\mathbf 0;\boldsymbol\mu_1,\boldsymbol\Sigma_1)
\mathcal{CN}(\mathbf 0;\boldsymbol\mu_2,\boldsymbol\Sigma_2)}
{\mathcal{CN}\left(\mathbf 0;
(\boldsymbol\Sigma_1^{-1}+\boldsymbol\Sigma_2^{-1})^{-1}
\left(\boldsymbol\Sigma_1^{-1}\boldsymbol\mu_1+
\boldsymbol\Sigma_2^{-1}\boldsymbol\mu_2\right),
(\boldsymbol\Sigma_1^{-1}+\boldsymbol\Sigma_2^{-1})^{-1}\right)}.
\end{align}

Moreover, if the terms $\bm\mu_1$, $\bm\Sigma_1$,
$\bm{\bar\mu}_2$, and $\bm{\bar\Sigma}_2$ are given,
we can derive the quotient of two N-dimensional complex Gaussian PDF
\begin{align}\label{Quotient}
&\frac{\mathcal{CN}(\mathbf x;\bm{\bar\mu}_2,\bm{\bar\Sigma}_2)}{\mathcal{CN}(\mathbf x;\bm\mu_1,\bm\Sigma_1)}=
\mathcal{CN}\left(\mathbf x;\left(\bm{\bar\Sigma}_2^{-1}-\bm\Sigma_1^{-1}\right)^{-1}
\left(\bm{\bar\Sigma}_2^{-1}\bm{\bar\mu}_2-\bm\Sigma_1^{-1}\bm\mu_1\right),
\left(\bm{\bar\Sigma}_2^{-1}-\bm\Sigma_1^{-1}\right)^{-1}\right)\notag\\
&\kern 10pt \times\frac{\mathcal{CN}(\mathbf 0;\bm{\bar\mu}_2,\bm{\bar\Sigma}_2)}{
\mathcal{CN}(\mathbf 0;\bm\mu_1,\bm\Sigma_1)
\mathcal{CN}\left(\mathbf 0;\left(\bm{\bar\Sigma}_2^{-1}-\bm\Sigma_1^{-1}\right)^{-1}
\left(\bm{\bar\Sigma}_2^{-1}\bm{\bar\mu}_2-\bm\Sigma_1^{-1}\bm\mu_1\right),
\left(\bm{\bar\Sigma}_2^{-1}-\bm\Sigma_1^{-1}\right)^{-1}\right)}.
\end{align}

%\newpage
%
%\begin{align}
%\mathcal{CN}(\mathbf x;\bm\mu_1,\bm\Sigma_1)
%\mathcal{CN}(\mathbf x;\bm\mu_2,\bm\Sigma_2)=
%\text{K}\mathcal{CN}(\mathbf x;\bm{\bar\mu}_2,\bm{\bar\Sigma}_2)
%\end{align}
%
%
%
%\begin{align}
%\bm{\bar\Sigma}_2=(\bm\Sigma_1^{-1}+ \bm\Sigma_2^{-1})^{-1}
%\rightarrow\bm{\bar\Sigma}_2^{-1}-\bm\Sigma_1^{-1}=\bm\Sigma_2^{-1}
%\end{align}
%\begin{align}
%&\bm{\bar\mu}_2= \bm{\bar\Sigma}_2(\bm\Sigma_1\bm^{-1}\mu_1+\bm\Sigma_2^{-1}\bm\mu_2)\rightarrow
%(\bm\Sigma_1\bm^{-1}\mu_1+\bm\Sigma_2^{-1}\bm\mu_2)=\bm{\bar\Sigma}_2^{-1}\bm{\bar\mu}_2\notag\\
%&\rightarrow\bm\Sigma_2^{-1}\bm\mu_2=\bm{\bar\Sigma}_2^{-1}\bm{\bar\mu}_2-\bm\Sigma_1^{-1}\bm\mu_1
%\rightarrow \bm\mu_2=\left(\bm{\bar\Sigma}_2^{-1}-\bm\Sigma_1^{-1}\right)^{-1}
%\left(\bm{\bar\Sigma}_2^{-1}\bm{\bar\mu}_2-\bm\Sigma_1^{-1}\bm\mu_1\right)
%\end{align}
%
%\begin{align}
%\frac{\mathcal{CN}(\mathbf x;\bm{\bar\mu}_2,\bm{\bar\Sigma}_2)}{\mathcal{CN}(\mathbf x;\bm\mu_1,\bm\Sigma_1)}&=
%\mathcal{CN}\left(\mathbf x;\left(\bm{\bar\Sigma}_2^{-1}-\bm\Sigma_1^{-1}\right)^{-1}
%\left(\bm{\bar\Sigma}_2^{-1}\bm{\bar\mu}_2-\bm\Sigma_1^{-1}\bm\mu_1\right),
%\left(\bm{\bar\Sigma}_2^{-1}-\bm\Sigma_1^{-1}\right)^{-1}\right)\notag\\
%&\times\frac{\mathcal{CN}(\mathbf 0;\bm{\bar\mu}_2,\bm{\bar\Sigma}_2)}{
%\mathcal{CN}(\mathbf 0;\bm\mu_1,\bm\Sigma_1)
%\mathcal{CN}\left(\mathbf 0;\left(\bm{\bar\Sigma}_2^{-1}-\bm\Sigma_1^{-1}\right)^{-1}
%\left(\bm{\bar\Sigma}_2^{-1}\bm{\bar\mu}_2-\bm\Sigma_1^{-1}\bm\mu_1\right),
%\left(\bm{\bar\Sigma}_2^{-1}-\bm\Sigma_1^{-1}\right)^{-1}\right)}
%\end{align}

\end{document}